\documentclass[aps,prd,reprint,amsmath,amssymb,superscriptaddress,floatfix]{revtex4}
\usepackage{graphicx}
\usepackage{amsfonts}
\usepackage{amssymb}
\usepackage{rotating}
\usepackage{booktabs}
\usepackage{xcolor}
\usepackage{soul}
\usepackage{color}
\usepackage{slashed}
\usepackage{multirow}
\usepackage{makecell}
\usepackage{epsf}
\usepackage{ulem}
\usepackage{cancel}
\usepackage{color,bm}
\usepackage[colorinlistoftodos]{todonotes}

\usepackage[colorlinks=true,citecolor=cyan,urlcolor=blue,bookmarks=true,bookmarks=true,bookmarksopen=true,bookmarksnumbered=true,bookmarksopenlevel=3]{hyperref}

\definecolor{airforceblue}{rgb}{0.36, 0.54, 0.66}
\definecolor{steelblue}{rgb}{0.27, 0.51, 0.71}
\definecolor{amber}{rgb}{1.0, 0.49, 0.0}

\begin{document}

\title{ Kotzinian-Mulders effect in semi-inclusive DIS within TMD factorization }
\author{Xuan Luo}
\author{\textsc{Hao Sun}\footnote{Corresponding author: haosun@mail.ustc.edu.cn \hspace{0.2cm} haosun@dlut.edu.cn}}
\affiliation{ Institute of Theoretical Physics, School of Physics, Dalian University of Technology, \\ No.2 Linggong Road, Dalian, Liaoning, 116024, P.R.China }
\date{\today}

\begin{abstract}
In this paper we study the Kotzinian-Mulders effect of a single hadron production in semi-inclusive deep inelastic scattering (SIDIS) within the framework of transverse momentum dependent (TMD) factorization.
The asymmetry is contributed by the convolution of the Kotzinian-Mulders function $g_{1T}$ and the unpolarized fragmentation function $D_1$.
As a TMD distribution, the Kotzinian-Mulders function in the coordinate space in the perturbative region can be represented as the convolution of the $C$-coefficients and the corresponding collinear correlation function.
The Wandzura-Wilczek approximation is used to obtain this correlation function.
We perform a detailed phenomenological numerical analysis of the Kotzinian-Mulders effect in the SIDIS process within TMD factorization at the kinematics of the HERMES and COMPASS measurements.
It is found that the obtained $x_B$-, $z_h$- and $P_{h\perp}$-dependent Kotzinian-Mulders effect are basically consistent with the HERMES and COMPASS measurements.
\vspace{0.5cm}
\end{abstract}
\maketitle
\setcounter{footnote}{0}

\section{introduction}

It is still a frontier of hadronic physics research to investigate the internal structure of the nucleon.
Azimuthal asymmetries in semi-inclusive deep inelastic scattering (SIDIS) are key observables to probe the spin dependent substructure of the nucleon.
Measurements of azimuthal asymmetries are crucial to comprehend the transverse structure of the proton.
The collinear picture utilized for DIS is not appropriated enough to get a variety of asymmetries in SIDIS, and the transverse momentum of the active quark in a nucleon has to be added.
The transverse momentum dependent (TMD) factorization \cite{Ji:2004xq,Ji:2004wu} approach can realize this asymmetry description.
The inclusive cross section of SIDIS is written as a convolution of Transverse Momentum Dependent Partonic Distribution Functions (TMD-PDFs), Transverse Momentum Dependent Fragmentation Functions (TMD-FFs) and QCD partonic cross sections.
In other words, SIDIS cross section can get factorized into TMD-PDFs having the information of the active quark distributions with transverse momentum inside the parent proton and the TMD-FFs illustrating the hadronizations of the struck quarks into the detected hadrons.
The azimuthal asymmetries in SIDIS were studied in lots of experiments.
The early work released by the JLab, HERMES, CLAS and COMPASS collaborations on azimuthal asymmetries in SIDIS production of charged hadrons was provided in Refs.\cite{Huang:2011bc,Adolph:2012sn,Adolph:2012sp,Adolph:2014zba,Alekseev:2010dm,Adolph:2016vou,Airapetian:2012yg,Airapetian:2019mov,Adolph:2014pwc,Schnell:2019oxw,Adolph:2015hta,Kravchenko:2012aaa,Avakian:2010ae,Pappalardo:2011cu,Airapetian:1999tv,Airapetian:2001eg,Airapetian:2005jc,Alexakhin:2005iw,Airapetian:2004tw,Avakian:2005ps,Airapetian:2020zzo}.
For both experimental and theorical reviews see \cite{Aidala:2012mv,Perdekamp:2015vwa,Avakian:2019drf,Anselmino:2020vlp,Bastami:2018xqd}.

The various azimuthal asymmetries in SIDIS were investigated theoretically by a number of works (e.g. \cite{Kotzinian:2006dw,Barone:2009hw,Mao:2014fma}).
In general, the authors explore SIDIS process at twist-two level in the parton model with TMDs and TMD FFs.
Such processes can be described in terms of eight PDFs including six time reversal even and two time reversal odd PDFs.
Among the leading-twist TMDs, the Kotzinian-Mulders(KM) function $g_{1T}(x,\vec{k}_T^2)$ \cite{Kotzinian:1995cz} describing the probability of discovering a longitudinally polarized quark inside a transversely polarized nucleon is rarely considered so far.
The $g_{1T}(x,\vec{k}_T^2)$ is chiral-even and can be reached in SIDIS combined with the unpolarized fragmentation function (FF).
In practice, $g_{1T}(x,\vec{k}_T^2)$ combined with unpolarized FF $D_1$ can be accessed from double spin asymmetries (DSA) $A_{LT}^{\cos(\phi_h-\phi_S)}$ in SIDIS.
The reason is that both a longitudinally polarized beam and a transversely polarized target are necessary to the longitudinal polarization of the active quark. 
This DSA is usually referred to KM effect \cite{Kotzinian:1995cz}.

In Ref.\cite{Kotzinian:2006dw}, the authors study the KM effect in SIDIS without scale evolution. In this paper we perform a more detailed phenomenological analysis of the KM effect in SIDIS within TMD factorization and compare the results with the data from COMPASS and HERMES Collaboration \cite{Parsamyan:2010se,Pappalardo:2011cu,Airapetian:2020zzo}.
There have also been measurements for KM effect in SIDIS with a neutron target \cite{Parsamyan:2007ju,Huang:2011bc}.
TMD factorization has been applied in many works \cite{Echevarria:2014xaa,Wang:2018pmx,Wang:2018naw,Li:2019uhj,Luo:2020hki,Xue:2020xba} focusing on various asymmetries in Drell-Yan and SIDIS.
Basing on the previous works by Collins-Soper-Sterman (CSS) \cite{Collins:1981uk,Collins:1984kg}, the so-called transverse momentum dependent (TMD) evolution following from factorization theorems has been well boosted in recent years.
Similar phenomenological studies for asymmetries contributed by Sivers, Boer-Mulders and Collins functions are discussed within TMD factorization in both Drell-Yan and SIDIS. 
The energy scale evolution  is connected with the Sudakov form factor \cite{Collins:1984kg,Collins:2011zzd,Collins:1999dz} after solving the evolution equation, which can be split into a perturbatively computable part $S_{\rm pert}$ and a nonperturbative part $S_{\rm NP}$. 
To be precise, TMD evolution is carried out in coordinate $b$-space related by momentum space via a Fourier transformation.
The use of $b$-space simplifies the expressions of the cross sections into products of $b$ dependent TMDs in contrast to convolutions in momentum space.
Then the Sudakov evolution kernel comes to be non-perturbative at large separation distances $b$,whereas at small $b \ll 1/\Lambda_{\rm QCD}$ it is perturbative and can be worked out order by order in strong coupling constant $\alpha_s$.
One needs to perform a two dimensional Fourier transform to the physical $k_\perp$ space for the corresponding TMDs to calculate the measured cross sections.
The $b$ dependence of TMDs related to their collinear counterparts, such as collinear parton distribution functions, fragmentation functions or multiparton correlation functions can be calculated in perturbation theory.
Specifically, the KM function $g_{1T}$ in the coordinate space in the perturbative region can be represented as the convolution of the $C$-coefficients and the corresponding collinear correlation functions, $\tilde{g}(x)$.
In this paper, the perturbative Sudakov form factors are considered up to the next-to-leading order accuracy and we adopt the tree-level results of the $C$-coefficients since the $C$-coefficients for $g_{1T}$ still remain in the leading order.
The nonperturbative Sudakov form factors in the unpolarized differential cross section are taken from Ref.\cite{Scimemi:2019cmh} which follows the CSS formalism with the $b^*$-prescription.
We perform the TMD evolution for reaching the fragmentation function and $\tilde{g}(x)$ at a initial scale $\mu_b=c/b^*$ by a evolution package QCDNUM \cite{Botje:2010ay}.
Based on the above considerations, in this paper, we estimate the KM effect within the TMD factorization and compare the results with the HERMES and COMPASS measurements.

The paper is organized as follows. In Sec.\ref{secII} we review the basic framework of TMD evolution for accessing the KM-Mulders effect in the SIDIS process. In Sec.\ref{secIII} We present the numerical calculation of the KM effect for the underlying process at the kinematics of HERMES and COMPASS measurements, respectively.
The conclusion of the paper in given in Sec.\ref{secIV}.

\section{framework}\label{secII}

We mostly follow the framework paper \cite{Wang:2018pmx}, which studied the Sivers asymmetry of the Drell-Yan process within TMD factorization. 
We consider the single hadron production in SIDIS by exchanging a virtual photon $q_\mu = l_\mu-l_\mu'$ with invariant mass $Q^2=-q^2$
\begin{eqnarray} \label{eq1}
\begin{aligned}
l^{\to}(\ell)+p^\uparrow(P) \to l'(\ell')+h(P_h)+X(P_X)
\end{aligned}
\end{eqnarray}
where a longitudinal polarized lepton scatters off a transverse polarized target nucleon with polarization $S$ and momentum $P$.
Inside the target, the photon hits the active quark with momentum $k$ and then changes it to $p$.
We adopt the usual SIDIS variables \cite{Meng:1991da}:
\begin{eqnarray} \label{eq2}
\begin{aligned}
S_{eP}=(l+P)^2, \qquad x_B=\frac{Q^2}{2P \cdot q}, \qquad y=\frac{P \cdot q}{P \cdot l}=\frac{Q^2}{x_B S_{eP}}, \qquad z_h=\frac{P \cdot P_h}{P \cdot q}, \qquad \gamma=\frac{2M x}{Q} 
\end{aligned}
\end{eqnarray}
When $P_{h\perp} \ll Q$, the TMD factorization applies and the SIDIS differential cross section including $\cos(\phi_h-\phi_S)$ term can be written as \cite{Bacchetta:2006tn}
\begin{eqnarray} \label{eq3}
\begin{aligned}
\frac{d^5 \sigma}{dx_B dy dz_h d^2 \vec{P}_{h\perp}}& = \sigma_0 \Bigg[F_{UU,T}+\sqrt{1-\varepsilon^2}\cos(\phi_h-\phi_S)F_{LT}^{\cos(\phi_h-\phi_S)}+\cdots \Bigg]
\\& = \sigma_0 \Bigg[\mathcal{I}[f_1 D_1]+\sqrt{1-\varepsilon^2} \cos(\phi_h-\phi_S)\mathcal{I}\bigg[ \frac{\hat{\vec{h}} \cdot \vec{k}_T}{M} g_{1T} D_1 \bigg]+\cdots \Bigg] 
\end{aligned}
\end{eqnarray}
where $\displaystyle \varepsilon=\frac{1-y-\frac{1}{4}\gamma^2 y^2}{1-y+\frac{1}{2}y^2+\frac{1}{4}\gamma^2 y^2}$
and
\begin{eqnarray} \label{eq4}
\begin{aligned}
\sigma_0 = \frac{2\pi\alpha_{em}^2}{Q^2} \frac{1+(1-y)^2}{y},
\end{aligned}
\end{eqnarray}
and $\vec{P}_{h\perp}$ is the transverse momentum of the final state hadron with respect to the lepton plane. 
Here $F_{UU}$ is the spin-averaged structure function, and $F_{LT}$ is the spin dependent structure function contributing to the $\cos(\phi_h-\phi_S)$ azimuthal asymmetry.
The unit vector $\hat{\vec{h}}=\vec{P}_{h\perp}/|\vec{P}_{h\perp}|$. We have introduced $\phi_h$ and $\phi_S$ being the azimuthal angles of the transverse momentum vector of the final-state hadron and the transverse spin of the target. 
These angles are defined in the target rest frame with the $\hat{z}$ axis along the virtual-photon momentum and the $\hat{x}$ axis along the lepton transverse momentum, which follow the Trento Conventions \cite{Bacchetta:2004jz}.
We have only kept the terms we are interested in. We have also adopted the notation
\begin{eqnarray} \label{eq5}
\begin{aligned}
\mathcal{I}[\omega f D]=\sum_q e_q^2 \int d^2 \vec{p}_T d^2 \vec{k}_T \delta^{(2)}(\vec{p}_T-\vec{k}_T-\frac{\vec{P}_{h\perp}}{z_h}) \omega(\vec{p}_T,\vec{k}_T)f^q(x_B,p_T^2)D^{h/q}(z_h,k_T^2)
\end{aligned}
\end{eqnarray}
where $\vec{k}_T$ and $\vec{p}_T$ are the corresponding transverse momentum componment of $k,p$. 
$\omega(\vec{p}_T,\vec{k}_T)$ is an arbitrary function in terms of $\vec{k}_T$ and $\vec{p}_T$.
The second term in the r.h.s. of Eq.(\ref{eq3}) refers to a leading twist effect involving the coupling of the transversal helicity distribution $g_{1T}$ and the unpolarized fragmentation function $D_1$.
In SIDIS experiments the KM effect can be accessed by
\begin{eqnarray} \label{eq6}
\begin{aligned}
A_{LT}^{\cos(\phi_h-\phi_S)}=\frac{2\int d\phi_h d\phi_S \cos(\phi_h-\phi_S)d\sigma_{LT}}{\int d\phi_h d\phi_S d\sigma_{UU}}=\frac{\int \sigma_0 \sqrt{1-\varepsilon^2}\mathcal{I}[ \frac{\hat{\vec{h}} \cdot \vec{k}_T}{M} g_{1T} D_1 ]}{\int \sigma_0\mathcal{I}[ f_{1} D_1 ]}
\end{aligned}
\end{eqnarray} 

In order to obtain a more detailed analysis of the KM effect, we have to consider the scale evolution.
It is convenient to perform the scale evolution of the TMD PDFs and FFs in the coordinate space (b-space). There are two scale parameters named $\zeta_F $(or $\zeta_D$) and $\mu$ in a general TMD PDF. The corresponding evolution equations describe these scale dependences.
The $\zeta$ scale evolution is presented with the Collins-Soper (CS) equation \cite{Collins:1981uk}:
\begin{eqnarray} \label{eq7}
\begin{aligned}
\frac{\partial \ln \widetilde{f}_1^q (x_B,b;\zeta_F,\mu)}{\partial \ln \sqrt{\zeta_F}}=\frac{\partial \ln \widetilde{D}_1^{h/q} (z_h,b;\zeta_D,\mu)}{\partial \ln \sqrt{\zeta_D}}=\widetilde{K}(b,\mu)
\end{aligned}
\end{eqnarray}
where $\widetilde{K}(b,\mu)$ denotes the CS kernel.
The $\mu$ dependence originates from  renormalization group equations for $\widetilde{f}_1^q$, $\widetilde{D}_1^{h/q}$ and $\widetilde{K}$
\begin{eqnarray} \label{eq8}
\begin{aligned}
\frac{d\widetilde{K}(b,\mu)}{d\ln\mu}&=-\gamma_K(\alpha_s(\mu))
\\
\frac{d\ln\widetilde{f}_1^q(x_B,b;\zeta_F,\mu)}{d\ln\mu}&=\gamma_F (\alpha_s(\mu),\zeta_F/\mu^2)
\\
\frac{d\ln\widetilde{D}_1^{h/q}(z_h,b;\zeta_D,\mu)}{d\ln\mu}&=\gamma_D (\alpha_s(\mu),\zeta_D/\mu^2)
\end{aligned}
\end{eqnarray}
where $\gamma_K$, $\gamma_F$ and $\gamma_D$ are anomalous dimensions of $\widetilde{K}$, $\widetilde{f}_1^q$ and $\widetilde{D}_1^{h/q}$, respectively.
On the ground of many previous discussion on solutions of above equations in Ref.\cite{Collins:1981uk,Collins:1984kg,Collins:2011zzd,Ji:2004wu,Collins:2014jpa,Ji:2004xq}, for numerical calculation we have to make a choice for values of $\zeta_F$ and $\zeta_D$. As stated in Ref.\cite{Aybat:2011zv}, we will treat the PDFs and FFs symmetrically and use $\sqrt{\zeta_F}=\sqrt{\zeta_D}=Q$.
Then we can express $f(x,b;\zeta_F=Q^2,\mu=Q)$ as $f(x,b,Q)$ for simplicity.we can summarize that the energy evolution of TMDs $(\widetilde{f})$ from a initial energy $\mu$ to another energy $Q$ can be represented by the Sudakov form factor in the exponential form $\exp(-S)$
\begin{eqnarray} \label{eq9}
\begin{aligned}
\widetilde{f}(x,b,Q)=\mathcal{F} \cdot e^{-S} \cdot \widetilde{f}(x,b,\mu)
\end{aligned}
\end{eqnarray}
where $\mathcal{F}$ is the hard factor depending on the scheme one chooses. For fragmentation function we have a similar form including a hard factor $\mathcal{D}$. The coefficients $\mathcal{F}$ and $\mathcal{D}$ have been studied in details in Ref.\cite{Prokudin:2015ysa}.

We consider the evolution of TMD function $\widetilde{f}(x,k_\perp;Q)$ probed at a energy scale $Q$ and carrying the collinear momentum fraction $x$ and a transverse momentum $k_\perp$.
It is convenient to reach energy evolution in the coordinate space, thus we adopt the Fourier transform of $\widetilde{f}(x,k_\perp;Q)$ in the two-dimensional $b$ space listed as \cite{Wang:2018pmx}
\begin{eqnarray} \label{eq10}
\begin{aligned}
\widetilde{f}(x,b;Q)=\int d^2 k_\perp e^{-ik_\perp \cdot b} f(x,k_\perp;Q)
\end{aligned}
\end{eqnarray}

In this paper we employ the Collins-Soper-Sterman(CSS) formalsim and pick an initial scale $Q_i=c/b$ for energy evolution. Here $c=2e^{-\gamma_E}$, and $\gamma_E \approx 0.577$ is the Euler's constant.
The energy evolution of TMD in the $b$-space from an initial scale $Q_i$ up to the scale $Q_f=Q$ is represented by \cite{Collins:2011zzd,Aybat:2011zv,Aybat:2011ge,Echevarria:2012pw}
\begin{eqnarray} \label{eq11}
\begin{aligned}
\widetilde{f}(x,b;Q)=\widetilde{f}(x,b;c/b)\exp\left\{ -\int_{c/b}^Q \frac{d\mu}{\mu} \left( a_1\ln\frac{Q^2}{\mu^2}+b_1 \right) \right\}\left( \frac{Q^2}{(c/b)^2} \right)^{-\widetilde{K}}
\end{aligned}
\end{eqnarray}
The coefficients $a_1$,$a_2$ and $\widetilde{K}$ can be expanded as a $\alpha_s/\pi$ series
\begin{eqnarray} \label{eq12}
\begin{aligned}
&a_1=\sum_{n=1}^\infty a_1^{(n)}\left(\frac{\alpha_s}{\pi}\right)^n
\\
&b_1=\sum_{n=1}^\infty b_1^{(n)}\left(\frac{\alpha_s}{\pi}\right)^n
\\
&\widetilde{K}=\sum_{n=1}^\infty \widetilde{K}^{(n)}\left(\frac{\alpha_s}{\pi}\right)^n
\end{aligned}
\end{eqnarray}
In our calculation, we will take $a_1^{(1)}$, $a_1^{(2)}$ and $b_1^{(1)}$ for the NLL accuracy:
\begin{eqnarray} \label{eq13}
\begin{aligned}
&a_1^{(1)}=C_F
\\
&a_1^{(2)}=\frac{C_F}{2}\left[ C_A \left(\frac{67}{18}-\frac{\pi^2}{6}\right)-\frac{10}{9} T_R n_f \right]
\\
&b_1^{(1)}=-\frac{3}{2}C_F
\\
&\widetilde{K}^{(1)}=0
\end{aligned}
\end{eqnarray}
where $C_F=4/3$, $C_A=3$ and $T_R=1/2$ are color factors.
$n_f$ is the the quark-antiquark active number of flavours into which the gluon may split. Its value depends on $Q$ and at the HERMES kinematics it can be definitely lower than five. We take $n_f=4$ in this work.
Fourier transforming back in transverse momentum space \cite{Wang:2018pmx},
\begin{eqnarray} \label{eq14}
\begin{aligned}
\widetilde{f}(x,k_\perp;Q)=\int \frac{d^2 b}{(2\pi)^2} e^{ik_\perp \cdot b} \widetilde{f}(x,b;Q)=\frac{1}{2\pi}\int_0^\infty db b J_0(k_\perp b) \widetilde{f}(x,b;Q)
\end{aligned}
\end{eqnarray}
where $J_0$ is the Bessel function of the zeroth order.
We should obtain the details of the whole $b \in [0,\infty]$ region, i.e. we have to extrapolate to the non-perturbative large-$b$ region.  
A non-perturbative Sudakov factor $R_{\rm NP}(x,b;Q)=\exp(-S_{\rm NP})$ is introduced by
\begin{eqnarray} \label{eq15}
\begin{aligned}
\widetilde{f}(x,b;Q)=\widetilde{f}_{\rm pert}(x,b_*;Q) R_{\rm NP}(x,b;Q)
\end{aligned}
\end{eqnarray}
where the perturbative part of the TMD $\widetilde{f}(x,b_*;Q)$ comes to be
\begin{eqnarray} \label{eq16}
\begin{aligned}
\widetilde{f}_{\rm pert}(x,b_*;Q)=\widetilde{f}\left(x,b;\frac{c}{b_*}\right)e^{-S_{\rm pert}(Q;b_*)}
\end{aligned}
\end{eqnarray}
The $b_*$ satisfies $b_*=b/\sqrt{1+(b/b_{max})^2}$. It has the property that $b_* \approx b$ at low values of $b$ and $b_* \approx b_{max}$ at the large $b$ values.
The typical value of $b_{max}$ is chosen about 1 GeV$^{-1}$ so that $b_*$ is always in the perturbative region.
This $b_*$-prescription introduces a cut-off value $b_{max}$ and allows for a smooth transition from perturbative region and avoids the Landau pole singularity in $\alpha_s$.
Then the total Sudakov-like form factor can be written as the sum of perturbatively calculable part and non-perturbative contribution
\begin{eqnarray} \label{eq17}
\begin{aligned}
S(Q;b)=S_{\rm pert}(Q;b_*)+S_{\rm NP}(Q;b)
\end{aligned}
\end{eqnarray}
and the perturbative part of the Sudakov form factor can be written as
\begin{eqnarray} \label{eq18}
\begin{aligned}
S_{\rm pert}(Q;b_*)=\int_{\mu_b}^Q \frac{d\mu}{\mu} \left[ A\ln\frac{Q^2}{\mu^2}+B \right],
\end{aligned}
\end{eqnarray}
where $\mu_b=c/b_*$.
In the region where $1/b \gg \Lambda_{QCD}$, the TMD PDF(FF) at a fixed scale in $b$-space can be expanded as the convolution of perturbatively calculable hard coefficients and the corresponding collinear PDFs(FFs) \cite{Collins:1981uk,Bacchetta:2013pqa}
\begin{eqnarray} \label{eq19}
\begin{aligned}
\widetilde{f}_{q/H}(x,b;\mu)=\sum_i C_{q \leftarrow i} \otimes f^{i/H}(x,\mu)
\\
\widetilde{D}_{H/q}(z,b;\mu)=\sum_j \frac{1}{z^2} \hat{C}_{j \leftarrow q} \otimes D^{H/j}(z,\mu)
\end{aligned}
\end{eqnarray}
where $\otimes$ appears for the convolution in the momentum fraction $x$($z$)
\begin{eqnarray} \label{eq20}
\begin{aligned}
C_{q \leftarrow i} \otimes f^{i/H}(x_B,\mu_b) \equiv \int_{x_B}^1 \frac{d\xi}{\xi} C_{q \leftarrow i}\left( \frac{x_B}{\xi},b;\mu_b,\zeta_F \right)f^{i/H}(\xi,\mu_b)
\\
\hat{C}_{j \leftarrow q} \otimes D^{H/j}(z_h,\mu_b) \equiv \int_{z_h}^1 \frac{d\xi}{\xi} \hat{C}_{j \leftarrow q}\left( \frac{z}{\xi},b;\mu_b,\zeta_F \right)D^{H/j}(\xi,\mu_b)
\end{aligned}
\end{eqnarray}
Therefore including the TMD evolution, TMDs can be expressed as 
\begin{eqnarray} \label{eq21}
\begin{aligned}
&\widetilde{f}_1^q (x_B,b;Q^2)=e^{-S_{\rm pert}(Q,b_*)-S_{\rm NP}^{f_1}(Q,b)}\widetilde{\mathcal{F}}_q \sum_i C_{q \leftarrow i} \otimes f_1^i(x_B,\mu_b)
\\
&\widetilde{D}_1^q (z_h,b;Q^2)=e^{-S_{\rm pert}(Q,b_*)-S_{\rm NP}^{D_1}(Q,b)}\frac{1}{z_h^2} \widetilde{\mathcal{D}}_q \sum_j \hat{C}_{j \leftarrow q} \otimes D_1^{h/j}(z_h,\mu_b)
\end{aligned}
\end{eqnarray}
The hard coefficients $C_i$, $\mathcal{F}$ for $f_1$ and $\hat{C}_j$, $\mathcal{D}$ for $D_1$ have been calculated up to NLO, while those for the transversal helicity distribution are still remained in leading order. Thus in this work we adopt for consistency the LO results of the $C$ coefficients for PDFs and FFs under considerations. 

Then we can obtain the unpolarized PDF and FF in b space as
\begin{eqnarray} \label{eq22}
\begin{aligned}
& \widetilde{f}_1^q (x_B,b;Q^2)=e^{-S_{\rm pert}(Q,b_*)-S_{\rm NP}^{f_1}(Q,b)} f_1^i(x_B,\mu_b)
\\
& \widetilde{D}_1^q (z_h,b;Q^2)=e^{-S_{\rm pert}(Q,b_*)-S_{\rm NP}^{D_1}(Q,b)} \frac{1}{z_h^2} D_1^{h/j}(z_h,\mu_b)
\end{aligned}
\end{eqnarray}
Thus we can obtain in the denominator of Eq.(\ref{eq6})
\begin{eqnarray} \label{eq23}
\begin{aligned}
\mathcal{I}[f_1^q D_1^q]=\sum_q e_q^2 \int_0^\infty \frac{bdb}{2\pi z_h^2}J_0\left(\frac{P_{h\perp }b}{z_h}\right) f_1^q(x_B,\mu_b)D_1^q(z_h,\mu_b)e^{-2S_{\rm pert}-S_{\rm NP} }
\end{aligned}
\end{eqnarray}

To obtain a more precise form for unpolarized PDF and FF, we follow a very recent work \cite{Scimemi:2019cmh} applying a exceptionally simple expression for the evolved TMD distributions
\begin{eqnarray}
\begin{aligned}
\widetilde{f}(x,b;Q)=\widetilde{f}(x,b;c/b)\left( \frac{Q^2}{(c/b)^2} \right)^{-\widetilde{K}}f_{\rm NP}(x,b)
\end{aligned}
\end{eqnarray}
we recall that this expression is same for unpolarized TMDPDF and TMDFF. The parameterizations of the non-perturbative functions $f_{\rm NP}$ and $D_{\rm NP}$(for FF) are
\begin{eqnarray} \label{eq25}
\begin{aligned}
&f_{\rm NP}(x,b)=\exp\left(-\frac{\lambda_1(1-x)+\lambda_2 x+x(1-x)\lambda_5}{\sqrt{1+\lambda_3 x^{\lambda_4} b^2}}b^2\right),
\\
&D_{\rm NP}(z,b)=\exp\left(-\frac{\eta_1 z	+\eta_2(1-z)}{\sqrt{1+\eta_3(b/z)^2}}\frac{b^2}{z^2}\right)\left( 1+\eta_4 \frac{b^2}{z^2} \right)
\end{aligned}
\end{eqnarray}
The anomalous dimensions are expressed in the following ansatz
\begin{eqnarray}
\begin{aligned}
\widetilde{K}\equiv \mathcal{D}(\mu,b)=\mathcal{D}_{\rm resum}(\mu,b^*(b))+c_0 b b^*(b)
\end{aligned}
\end{eqnarray}
where $b^*(b)=b/\sqrt{1+b^2/B_{\rm NP}^2}$. The function $\mathcal{D}_{\rm resum}$ is the resummed perturbative expansion of anomalous dimensions, At LO it reads
\begin{eqnarray}
\begin{aligned}
\mathcal{D}_{\rm resum}^{\rm LO}=-\frac{\Gamma_0}{2\beta_0}\ln(1-\beta_0 \alpha_s(\mu)L_\mu)
\end{aligned}
\end{eqnarray}
with $\Gamma_0=4C_F$, $\beta_0=\frac{11}{3}C_A-\frac{2}{3}n_f$, and $L_\mu=\ln\bigg( \mu^2 b^2/(4e^{-2\gamma_E}) \bigg)$.
We choose $B_{\rm NP},c_0,\lambda_i,\eta_i$ values fitted by \cite{Scimemi:2019cmh} using HERA20 PDF sets and DSS FF sets. 
Then similar to the procedure above, we can obtain
\begin{eqnarray} \label{eq24}
\begin{aligned}
\mathcal{I}[f_1^q D_1^q]=\sum_q e_q^2 \int_0^\infty \frac{bdb}{2\pi z_h^2}J_0\left(\frac{P_{h\perp }b}{z_h}\right) f_1^q(x_B,\mu_b)D_1^q(z_h,\mu_b)f_{\rm NP}(x_B,b)D_{\rm NP}(z_h,b)\left( \frac{Q^2}{(c/b)^2} \right)^{-2\mathcal{D}(\mu_b,b)}
\end{aligned}
\end{eqnarray}

Now, we turn to the $\cos(\phi_h-\phi_S)$ asymmetry in SIDIS. 
In the small b region, we can also express the KM function $g_{1T}$ of the nucleon at a
fixed energy scale $\mu$ in terms of the perturbatively calculable coefficients and the corresponding collinear correlation function \cite{Boer:2011xd}
\begin{eqnarray} \label{eq26}
\begin{aligned}
\widetilde{g}_{1T}^{\alpha,q}(x,b;\mu)=2M\left( \frac{ib_\perp^\alpha}{2} \right)\tilde{g}_{q}(x,\mu)
\end{aligned}
\end{eqnarray}
where $M$ is the mass of the nucleon. The hard coefficients are calculated up to LO, and the KM function in the b space is defined as
\begin{eqnarray} \label{eq27}
\begin{aligned}
\widetilde{g}_{1T}^{\alpha,q}(x,b;\mu)=\int d^2 \vec{k}_\perp e^{-i\vec{k}_\perp \cdot \vec{b}_\perp} \frac{k_\perp^\alpha}{M}g_{1T}^{q}(x,\vec{k}_\perp^2;\mu)
\end{aligned}
\end{eqnarray}
The collinear function $\tilde{g}_{q}(x)$ is a twist-3 quark-gluon-quark correlation function, which is just the first transverse moment of the $g_{1T}$ \cite{Zhou:2008mz}
\begin{eqnarray} \label{eq28}
\begin{aligned}
\tilde{g}_{q}(x)=\int d^2 \vec{k}_\perp \frac{\vec{k}_\perp^2}{2M^2}g_{1T}^{q}(x,\vec{k}_\perp^2)=g_{1T}^{q(1)}(x)
\end{aligned}
\end{eqnarray}
As for the nonperturbative part of the Sudakov form factor associated with the KM function, the information still remains unknown. In a practical calculation, we assume that it is the same as $S_{\rm NP}^{\rm Siv}$ reached from Ref.\cite{Echevarria:2014xaa}.
Therefore, we can obtain the KM function in b-space as
\begin{eqnarray} \label{eq29}
\begin{aligned}
\tilde{g}_{1T}^{\alpha,q}(x,b)=2M\left( \frac{ib_\perp^\alpha}{2} \right) e^{-S_{\rm pert}-S_{\rm NP}}\tilde{g}_{q}(x)
\end{aligned}
\end{eqnarray}
Thus we can write the numerator in Eq.(\ref{eq6}) as 
\begin{eqnarray} \label{eq30}
\begin{aligned}
\mathcal{I}\bigg[ \frac{\hat{\vec{h}} \cdot \vec{k}_T}{M} g_{1T} D_1 \bigg]=\frac{1}{2\pi z^2}\int_0^\infty db b^2 J_1\left( \frac{P_{h\perp}b}{z} \right)\sum_q e_q^2 \tilde{g}_q(x_B,\mu_b) D_1^q(z_h,\mu_b)e^{-(S_{\rm NP}^{\rm Siv}+S_{\rm NP}^{D_1}+2S_{\rm pert})}
\end{aligned}
\end{eqnarray}

\section{Numerical calculation} \label{secIII}

In this section, we will present predictions of the KM effect in SIDIS with a longitudinally polarized lepton scattering off a transversely polarized proton at the kinematics of HERMES and COMPASS experiments	.
To obtain the numerical estimate of the denominator of the effect presented in Eq.(\ref{eq23}), we employ the LO set for MSTW2008 parametrization \cite{Lai:2010vv} for the unpolarized distribution function $f_1(x)$ of the proton.
We use the NLO fit \cite{deFlorian:2014xna} for the unpolarized parton-to-pion fragmentation function since we apply the TMD evolution at NLL accuracy. Meanwhile, we adopt a recent NLO fit \cite{deFlorian:2017lwf} for the unpolarized parton-to-Kaon fragmentation function.
For the numerator of the effect given in Eq.(\ref{eq30}), we have to parameterize the distribution $\tilde{g}(x)$ in a properly initial scale $\mu$ and then evolve it to the scale $\mu_b=c/b^*$.
After this, the TMD evolution equations will be used to evolve from $c/b_*$ to $Q$.
Since $g_{1T}$ has not been extracted from experiment data, we reach $\tilde{g}(x)$ by employing the Wandzura-Wilczek approximation \cite{Bastami:2018xqd}
\begin{eqnarray} \label{eq31}
\begin{aligned}
\tilde{g}(x,\mu_0)=g_{1T}^{(1)q}(x,\mu_0) \ensuremath{\stackrel{\text{\scriptsize WW--type}}{\approx}} x\int_x^1 \frac{dy}{y} g_1^q(y,\mu_0)
\end{aligned}
\end{eqnarray}
where $g_1^q$ is the quark helicity distibution extracted form Ref.\cite{deFlorian:2009vb} and 
$\mu_0=1$GeV.

As for the scale evolution of the $\tilde{g}$, we assume at the initial scale $Q_0=1$GeV the $\tilde{g}$ function is parameterized as Eq.(\ref{eq31}) and then evolve it to the final scale $Q$ using the evolution equation for $\tilde{g}$.
The energy evolution of the $\tilde{g}$ function has been studied extensively in literture \cite{Zhou:2008mz}.
Following Ref.\cite{Wang:2018pmx} and Ref.\cite{Kang:2015msa}, where only the homogeneous terms of the evolution kernel are kept in order to reach the evolution of the Qiu-Sterman function and twist-3 fragmentation function $\hat{H}^{(3)}$, respectively.
In this paper, we keep the same choice	. 
Similar choice was adopted as well in Ref.\cite{Luo:2020hki} studing the Sivers asymmetry in SIDIS. 
According to Eq.22 of \cite{Zhou:2008mz}, the homogenous terms of the $\tilde{g}$ evolution kernel are written as 
\begin{eqnarray} \label{eq32}
\begin{aligned}
P_{qq}^{\tilde{g}} \approx C_F\left[ \frac{1+z^2}{(1-z)_+}+\frac{3}{2}\delta(1-z) \right]-\frac{C_A}{2}\frac{1+z^2}{1-z}
\end{aligned}
\end{eqnarray}

\begin{figure}[htp]
\centering
\includegraphics[scale=0.66]{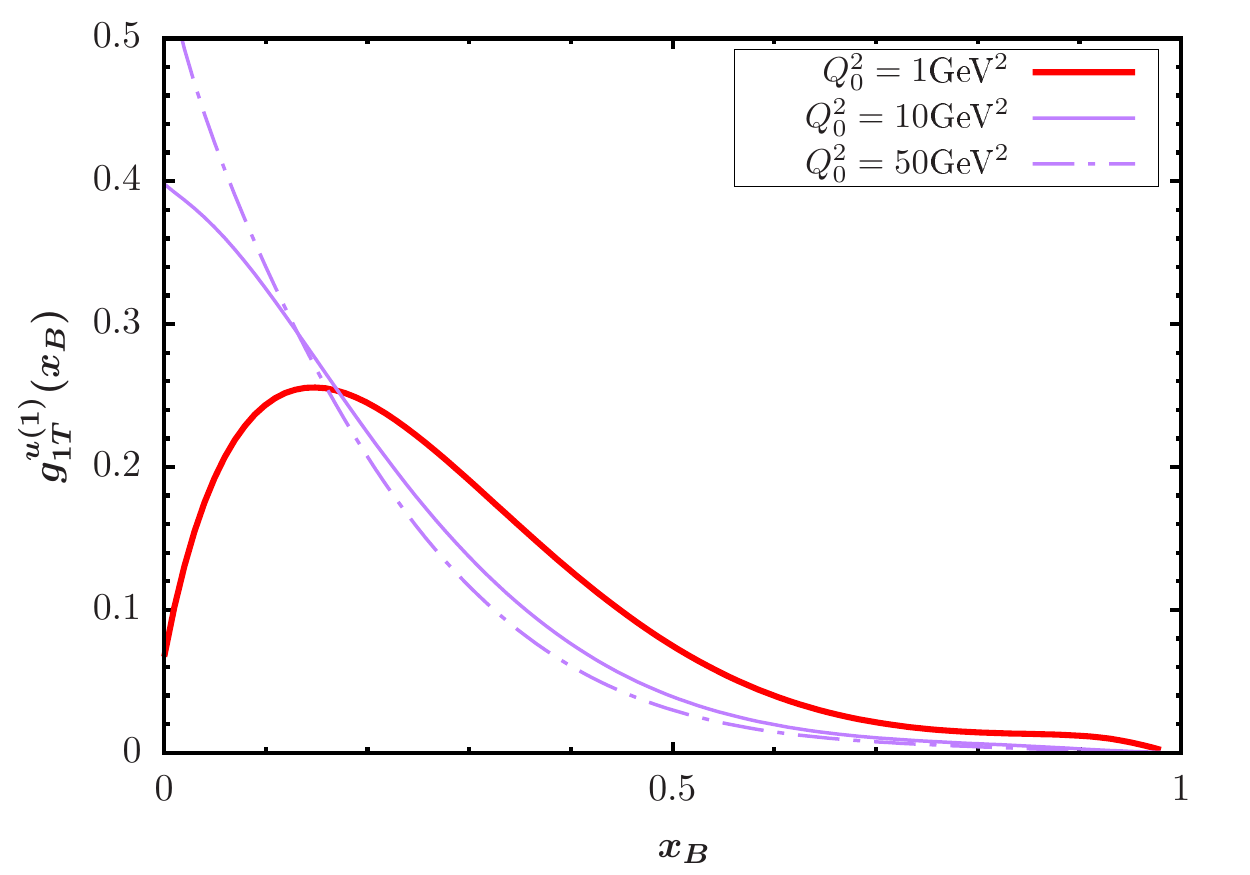}
\includegraphics[scale=0.66]{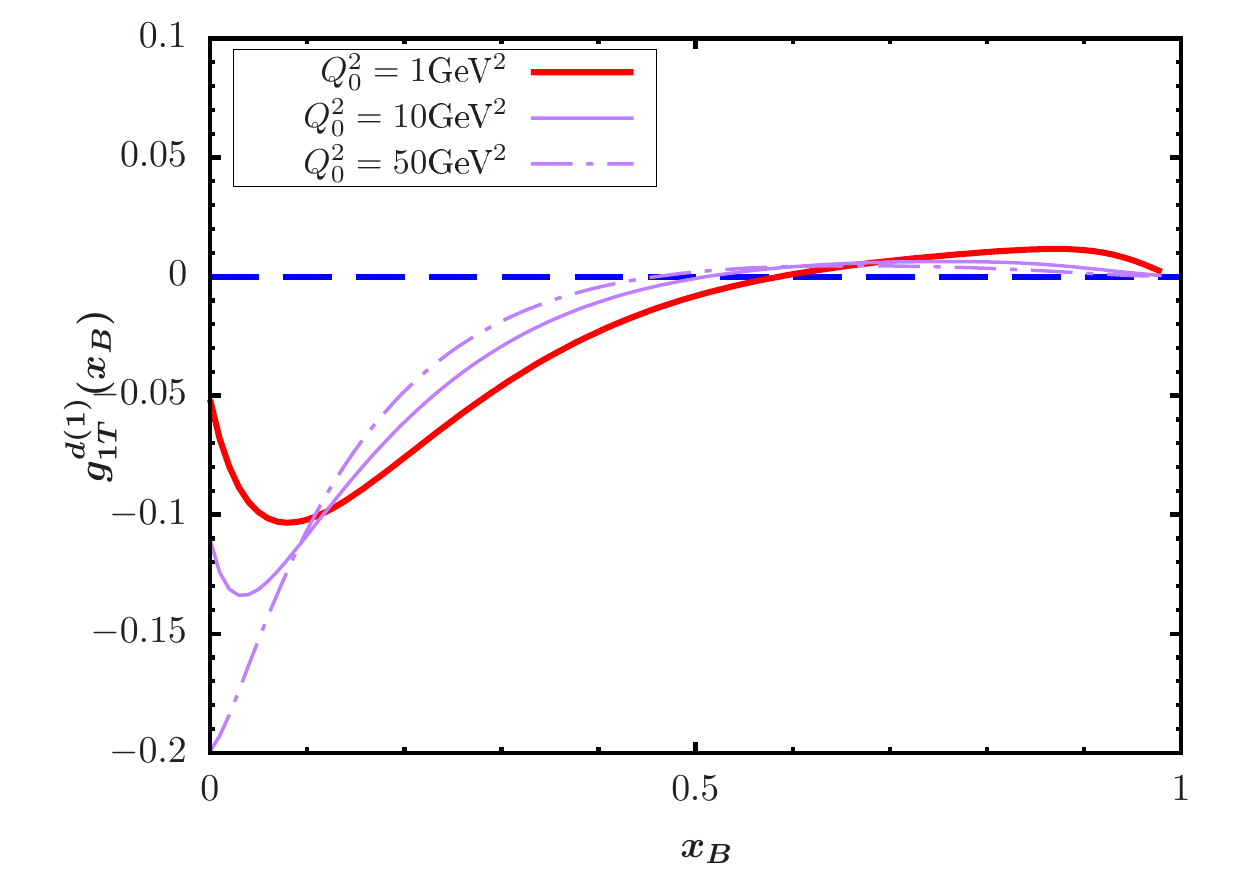}
\caption{ The $g_{1T}^{(1)}$ distributions of up and down quark as a function of $x_B$. }
\label{fig0}
\end{figure}

Numerical solution of $\tilde{g}(x)$'s evolution equation is performed by QCDNUM evolution package \cite{Botje:2010ay}. The energy evolution of fragmentation function is performed by the internal time-like evolution in QCDNUM.
The two loop QCD coupling constant \cite{Prosperi:2006hx} has been used in the evolution package and CSS evolution.
Original code of QCDNUM is modified by us so that $\tilde{g}(x)$ function evolution kernel is added, the initial scale for the evolution is chosen to be $Q_0^2=1$GeV$^2$. The QCDNUM code is executed with $\alpha_s(Q_0)=0.327$.
In Fig.\ref{fig0} we plot the $g_{1T}^{(1)}$ distributions of up and down quark at three scales. In both panel, the red solid lines depict the results at the initial scale $Q_0^2=1$GeV$^2$, and the purple solid and dotted lines show the results at $Q_0^2=10$GeV$^2$ and $Q_0^2=50$GeV$^2$ after applying the evolution equation for $\tilde{g}$. We find that the evolution equation significantly changes the shape and size of the TMD at different scale. The absolute maximum of TMD becomes large as the scale increases in both up and down quark cases.
The first analysis of the $g_{1T}^{(1)}$(defined as $\frac{1}{2}\sum_q e_q^2 g_{1T}^{(q(1))}(x)$) distributions was made in Fig.2 of \cite{Kotzinian:1995cz} where the authors employed the Wandzura-Wilczek approximation \cite{Bastami:2018xqd} by applying the BBS-parameterizations \cite{Brodsky:1994kg} for $g_1$. For comparsion, we have calculated the $g_{1T}^{(1)}$ by considering only $u$ and $d$ quark contributions with $g_1$ distribution from Ref.\cite{deFlorian:2009vb}. We find that two obtained $g_{1T}^{(1)}$ distributions are very consistent with each other both in shape and size. The only small difference is that the obtained $g_{1T}^{(1)}$ distribution by us is larger than that in \cite{Kotzinian:1995cz} very slightly.

To perform numerical calculations for $A_{LT}^{\cos(\phi_h-\phi_S)}$ in SIDIS at HERMES, we adopt the following kinematical cuts \cite{Airapetian:2009ae}
\begin{eqnarray} \label{eq47}
\begin{aligned}
&0.023 <x_B< 0.4 \qquad 0.1<y<0.95 \qquad 0.2<z_h<0.7 \qquad P_{h\perp} > 0.1\text{GeV}
\\
&Q^2>1\text{GeV}^2 \qquad W^2>10\text{GeV}^2
\end{aligned}
\end{eqnarray}
where $W$ is the invariant mass of photon-nucleon system with $W^2=(P+q)^2 \approx \frac{1-x_B}{x_B}Q^2$.
Furthermore, like Ref.\cite{Echevarria:2014xaa}, we choose $P_{h\perp} \le 0.6$GeV for hadron production at HERMES since we focus on the region $P_{h\perp} \le Q$ region where the TMD factorization applies.
At COMPASS, we choose \cite{Parsamyan:2010se}
\begin{eqnarray} \label{eq48}
\begin{aligned}
&0.004 <x_B< 0.7 \qquad 0.1<y<0.9 \qquad 0.2<z_h<1 \qquad 0.1<P_{h\perp}<0.6\text{GeV}
\\
&Q^2>1\text{GeV}^2 \qquad W^2>25\text{GeV}^2
\end{aligned}
\end{eqnarray}

\begin{figure}[htp]
\centering
\includegraphics[scale=0.46]{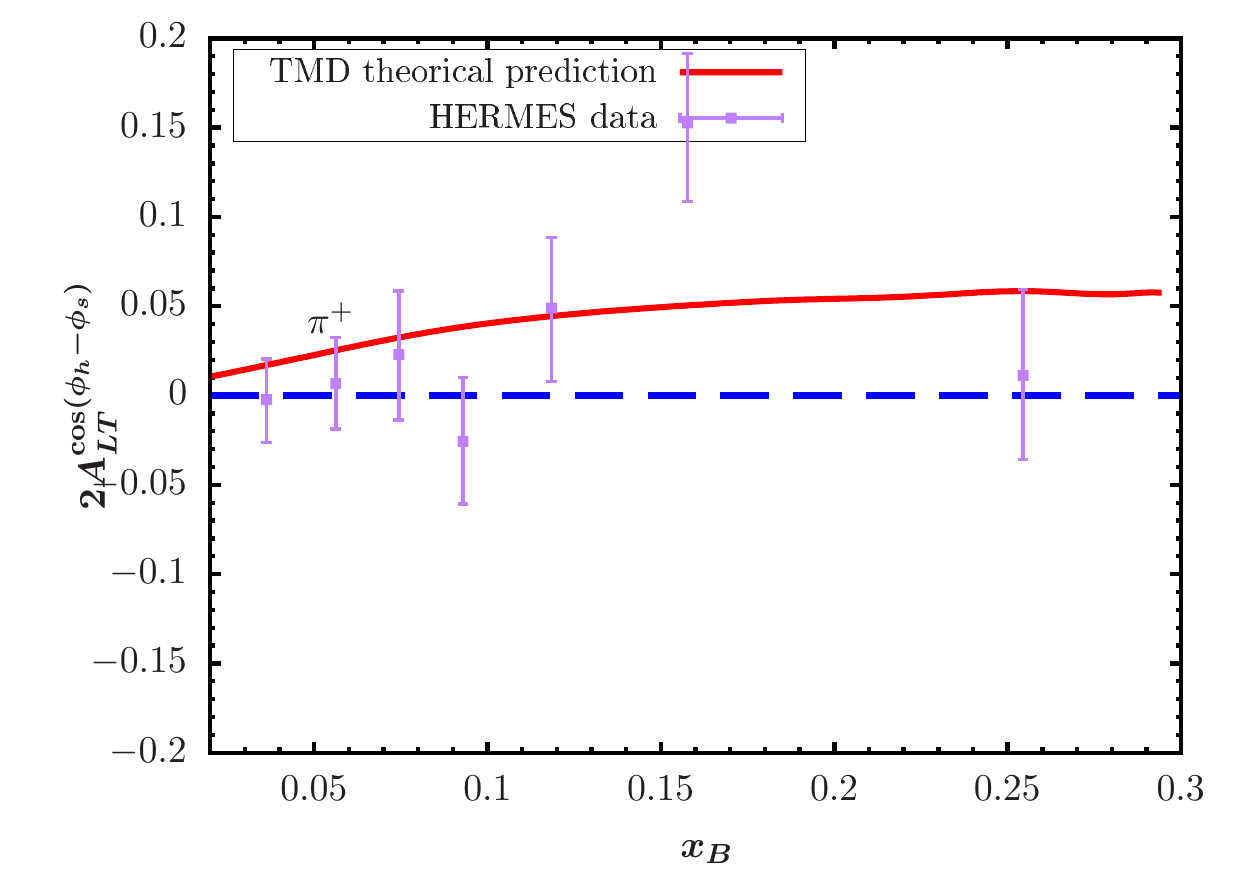}
\includegraphics[scale=0.46]{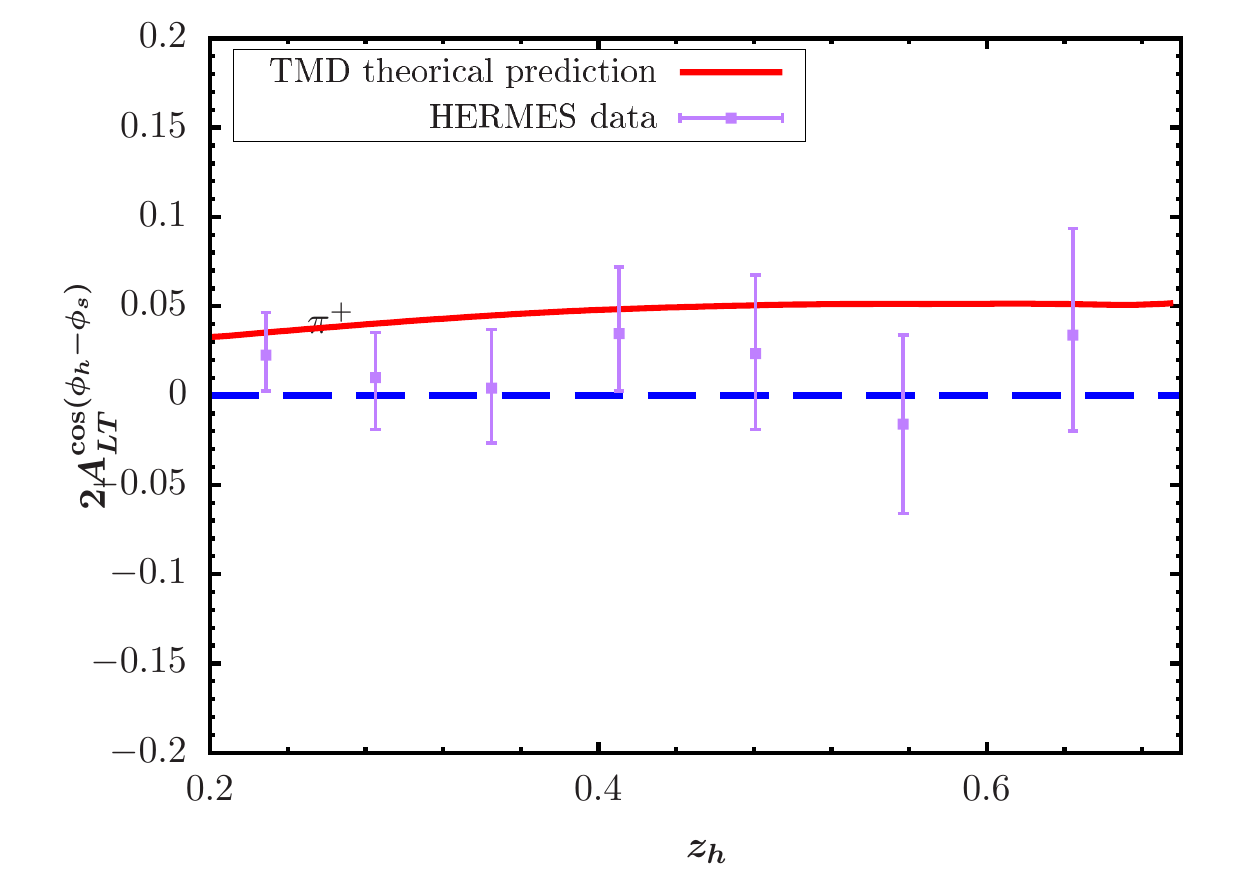}
\includegraphics[scale=0.46]{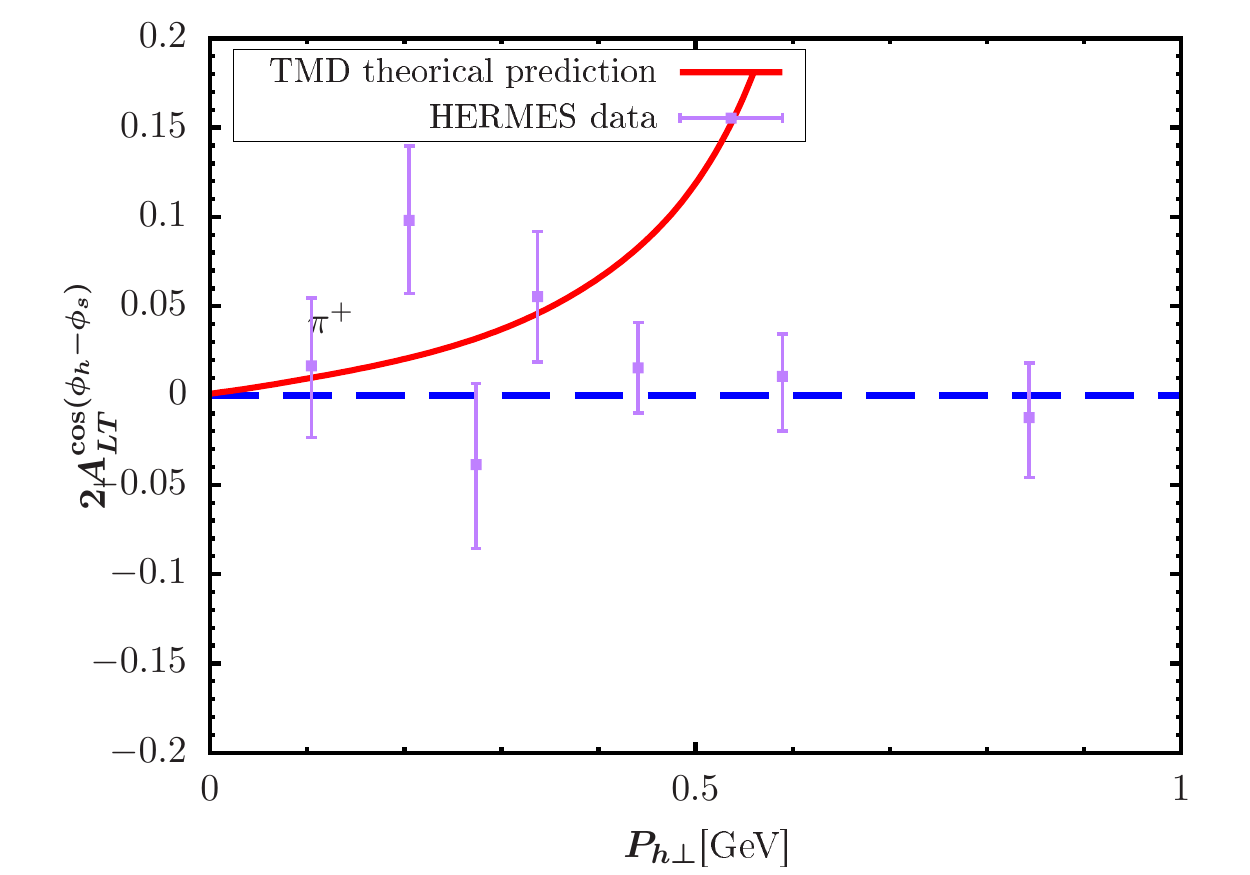}
\caption{ The KM effect calculated within TMD factorization, compared with the HERMES measurement \cite{Airapetian:2009ae} for $\pi^+$ production.}
\label{fig1}
\end{figure}

\begin{figure}[htp]
\centering
\includegraphics[scale=0.46]{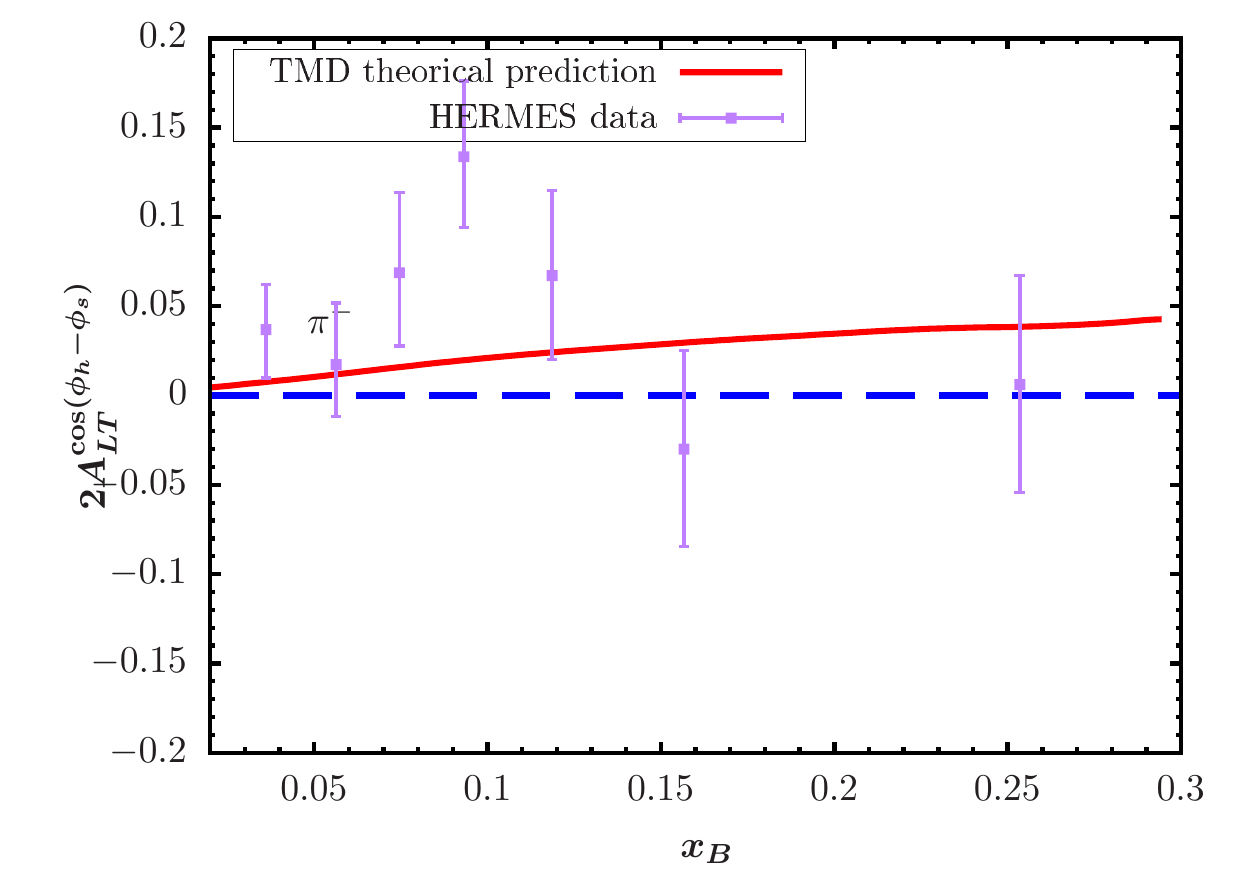}
\includegraphics[scale=0.46]{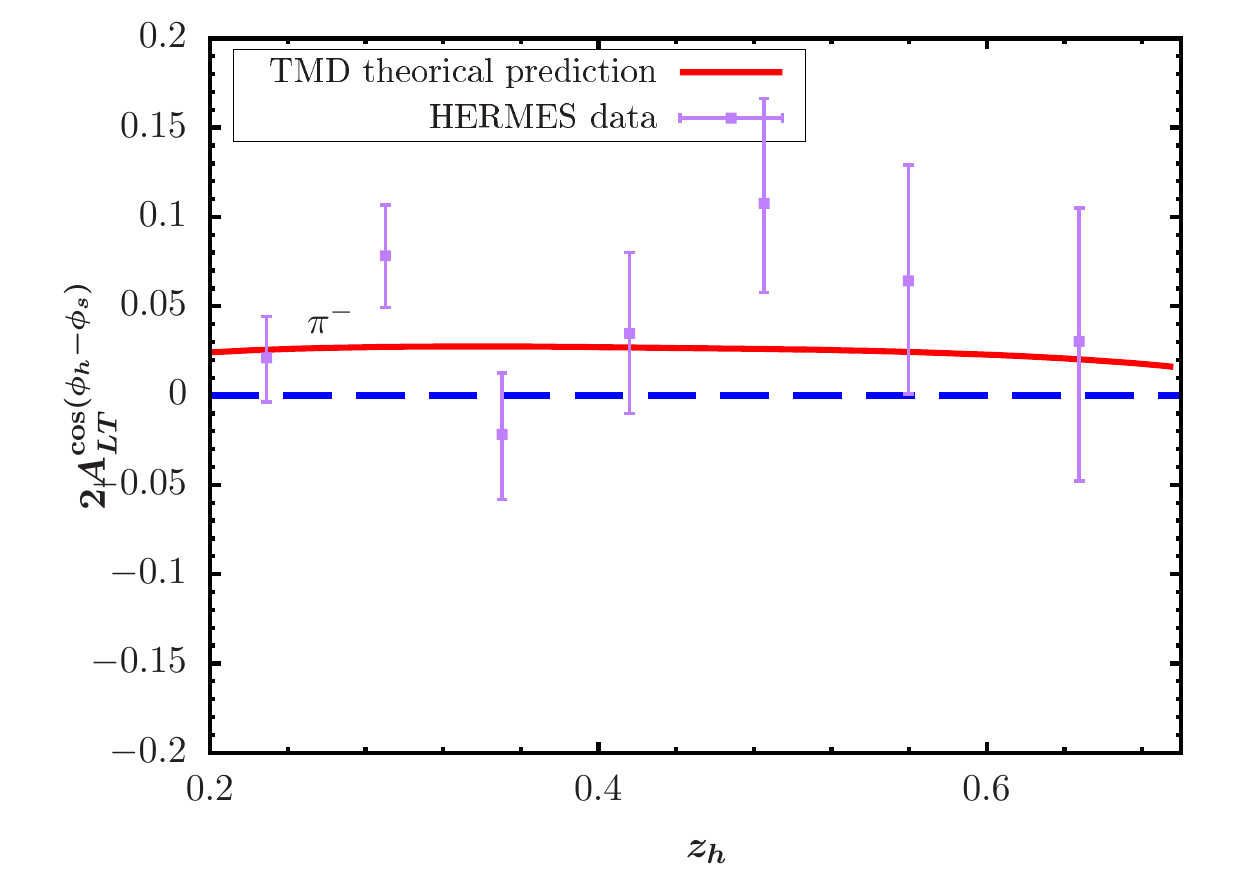}
\includegraphics[scale=0.46]{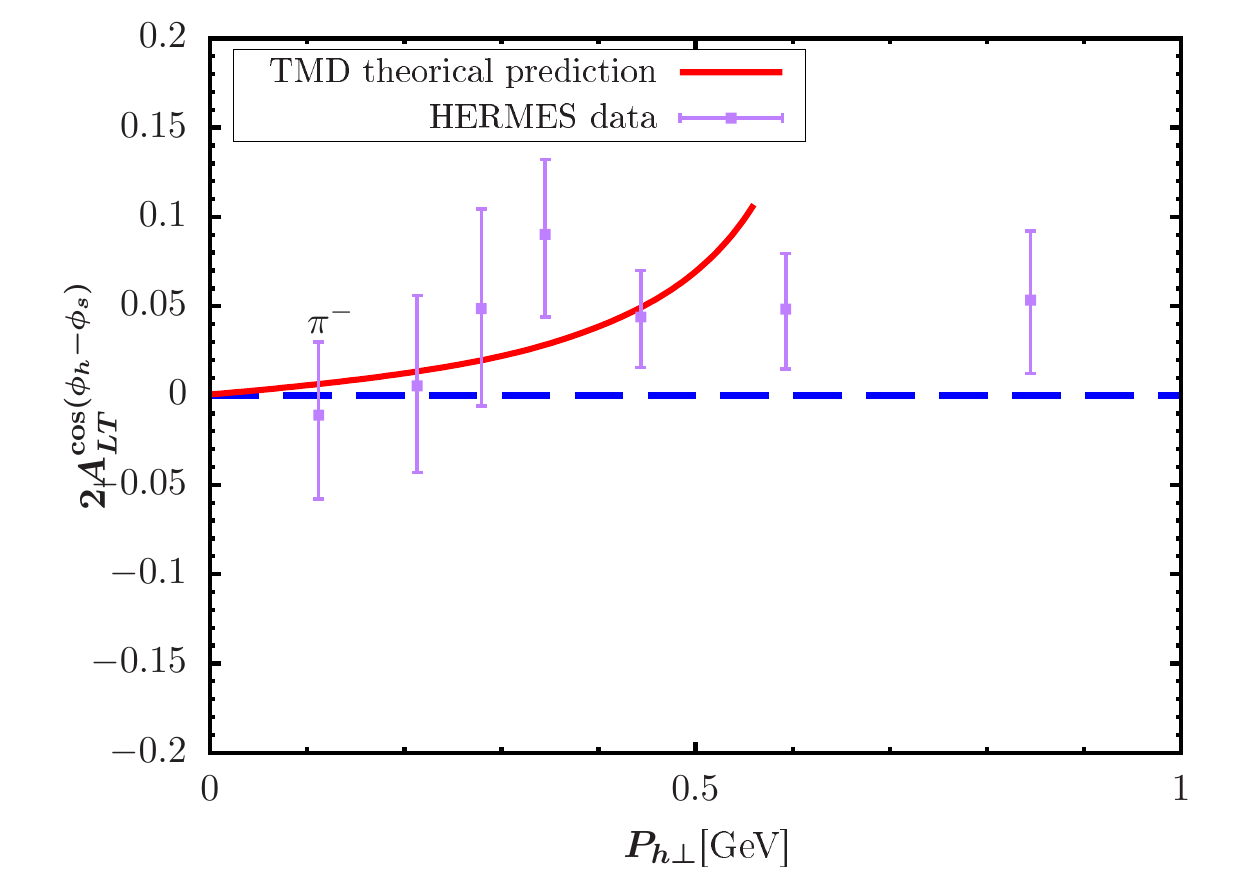}
\caption{ The KM effect calculated within TMD factorization, compared with the HERMES measurement \cite{Airapetian:2009ae} for $\pi^-$ production.}
\label{fig2}
\end{figure}

\begin{figure}[htp]
\centering
\includegraphics[scale=0.46]{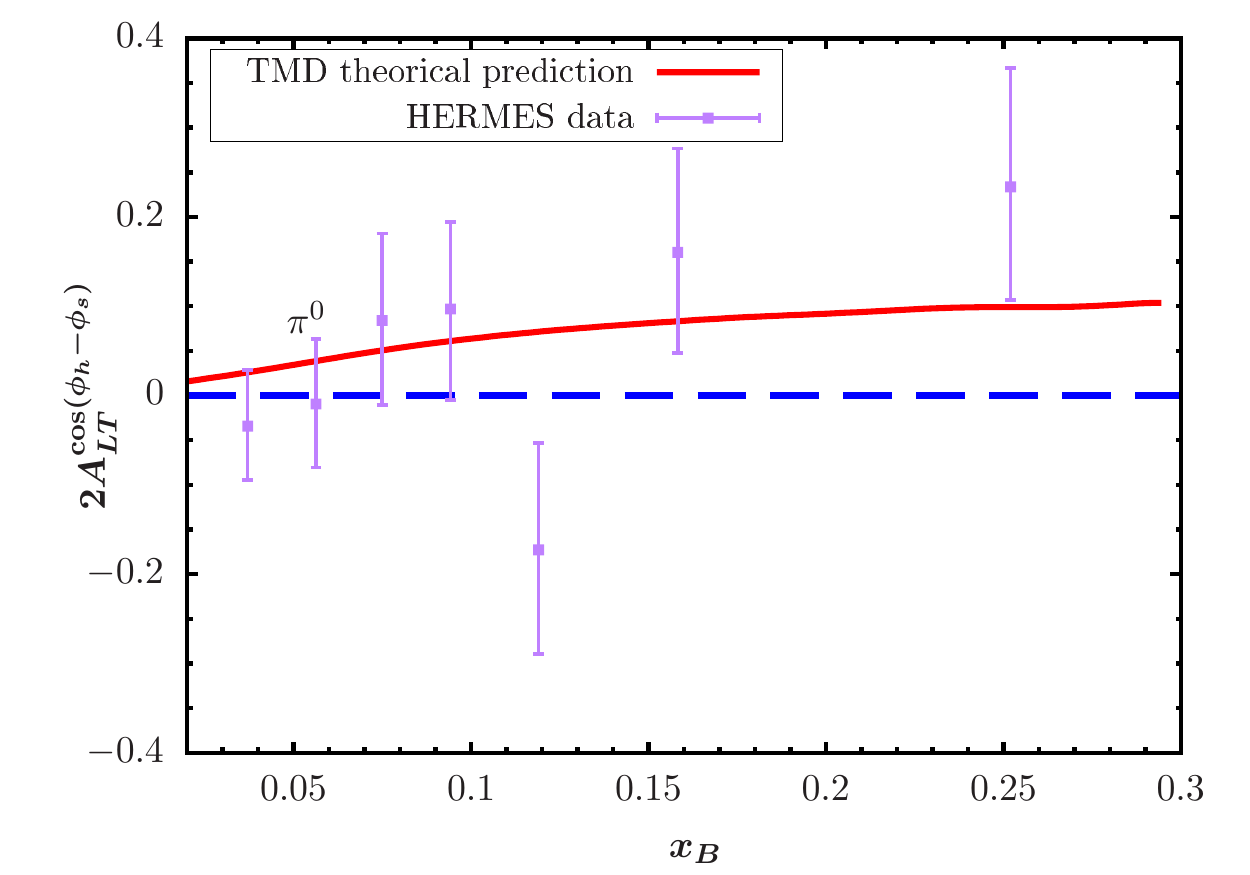}
\includegraphics[scale=0.46]{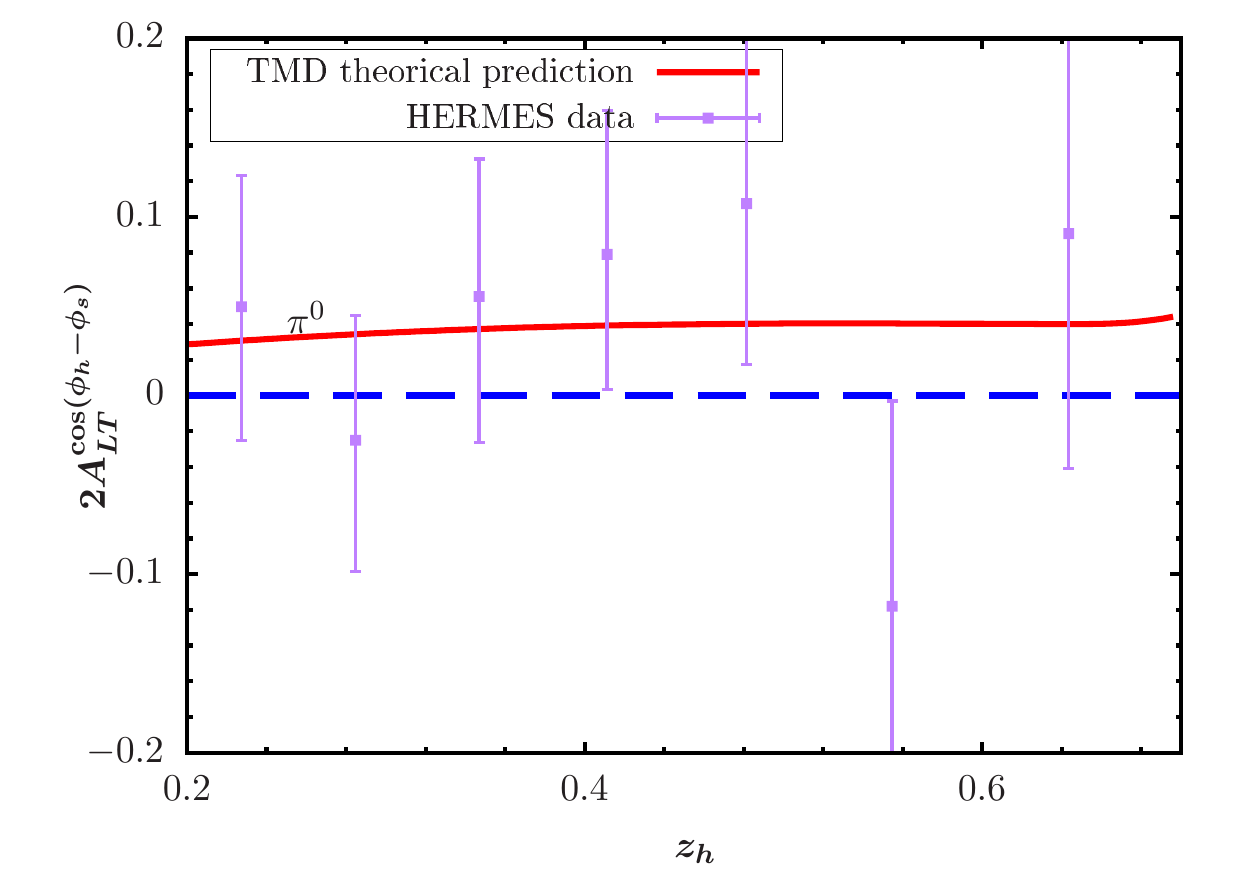}
\includegraphics[scale=0.46]{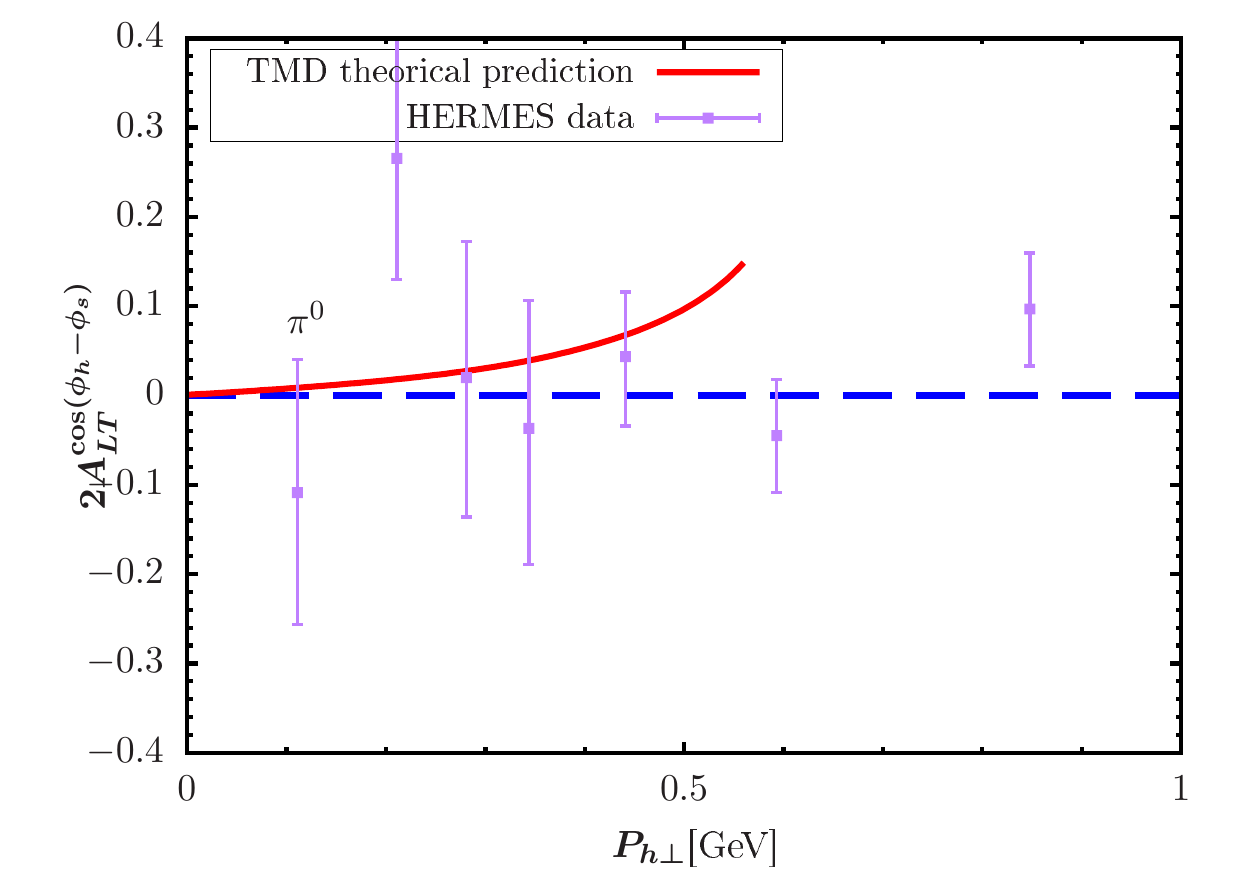}
\caption{ The KM effect calculated within TMD factorization, compared with the HERMES measurement \cite{Airapetian:2009ae} for $\pi^0$ production.}
\label{fig3}
\end{figure}

\begin{figure}[htp]
\centering
\includegraphics[scale=0.46]{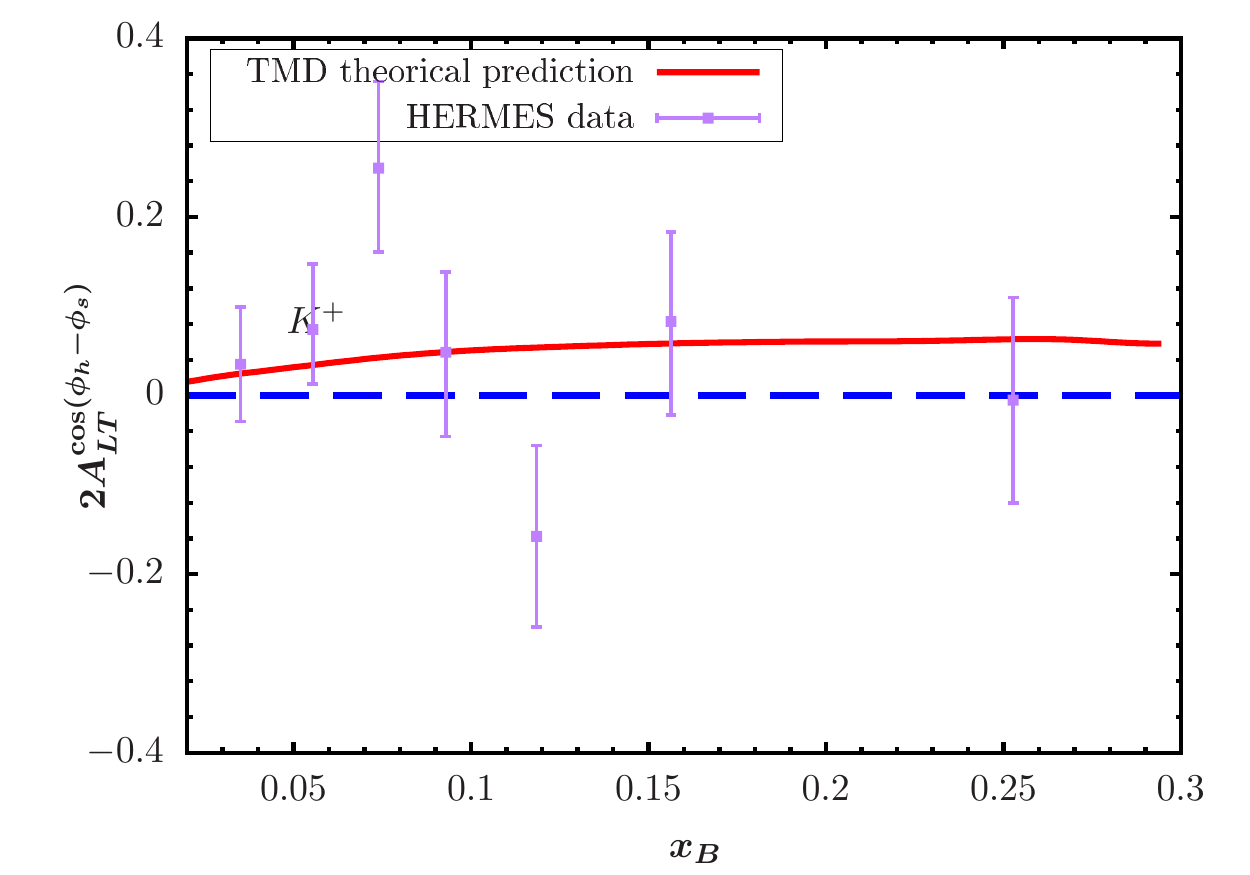}
\includegraphics[scale=0.46]{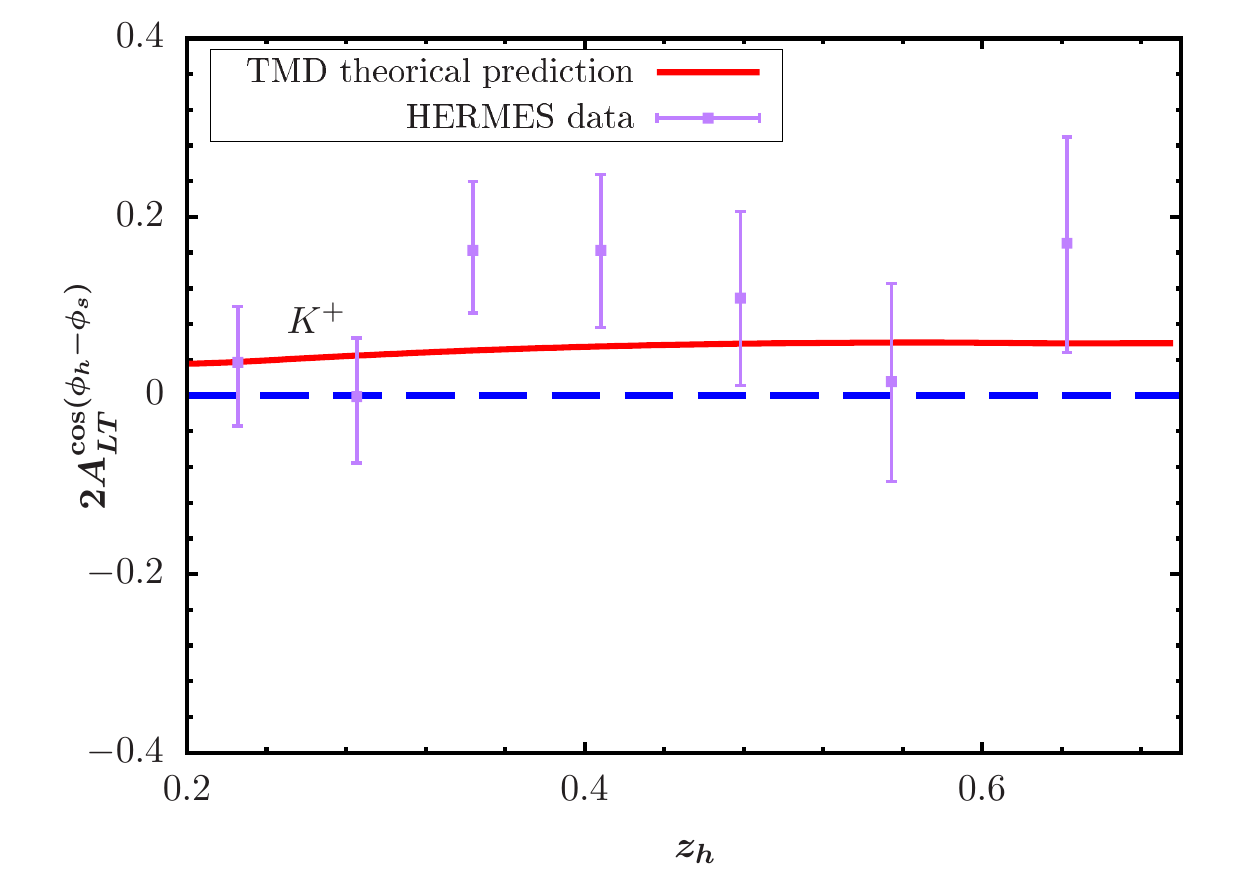}
\includegraphics[scale=0.46]{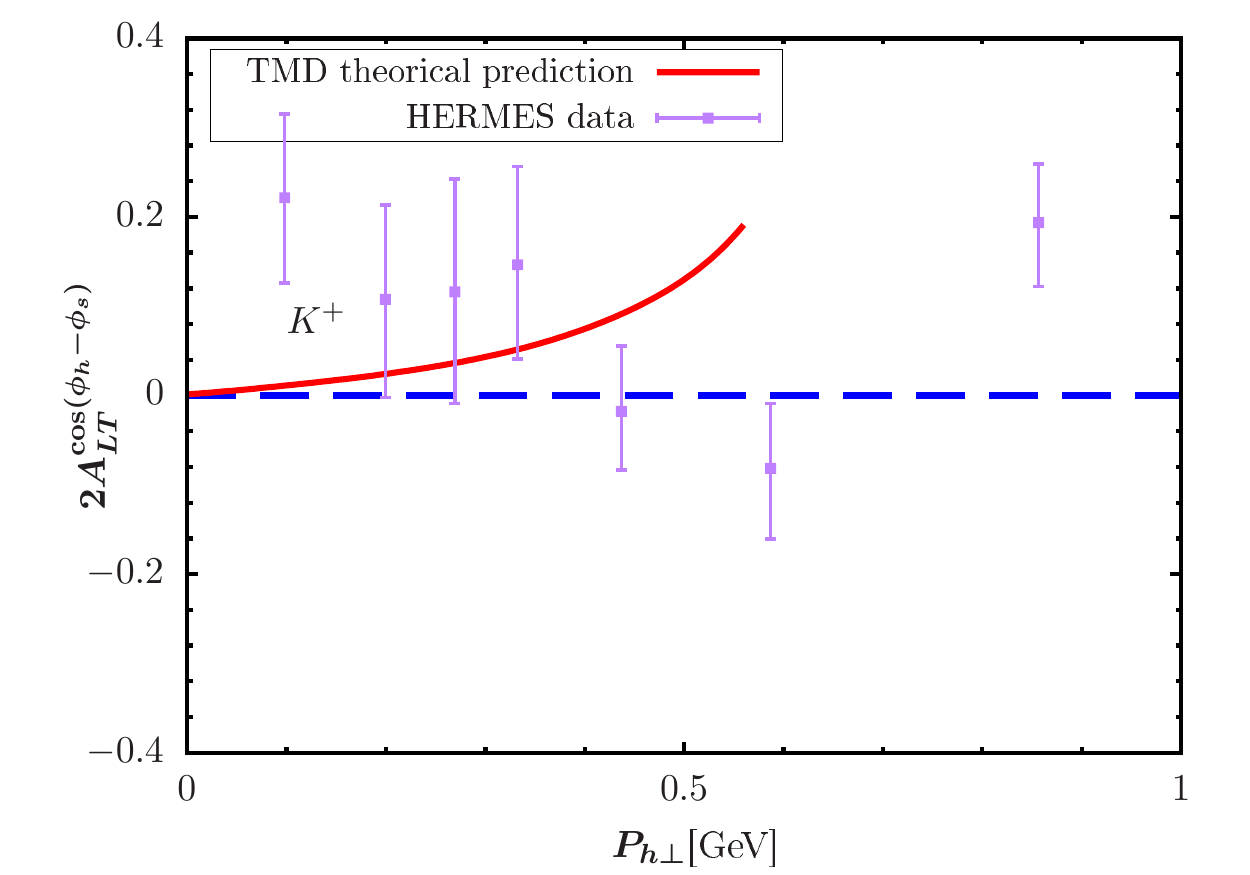}
\caption{ The KM effect calculated within TMD factorization, compared with the HERMES measurement \cite{Airapetian:2009ae} for $K^+$ production.}
\label{fig4}
\end{figure}

\begin{figure}[htp]
\centering
\includegraphics[scale=0.46]{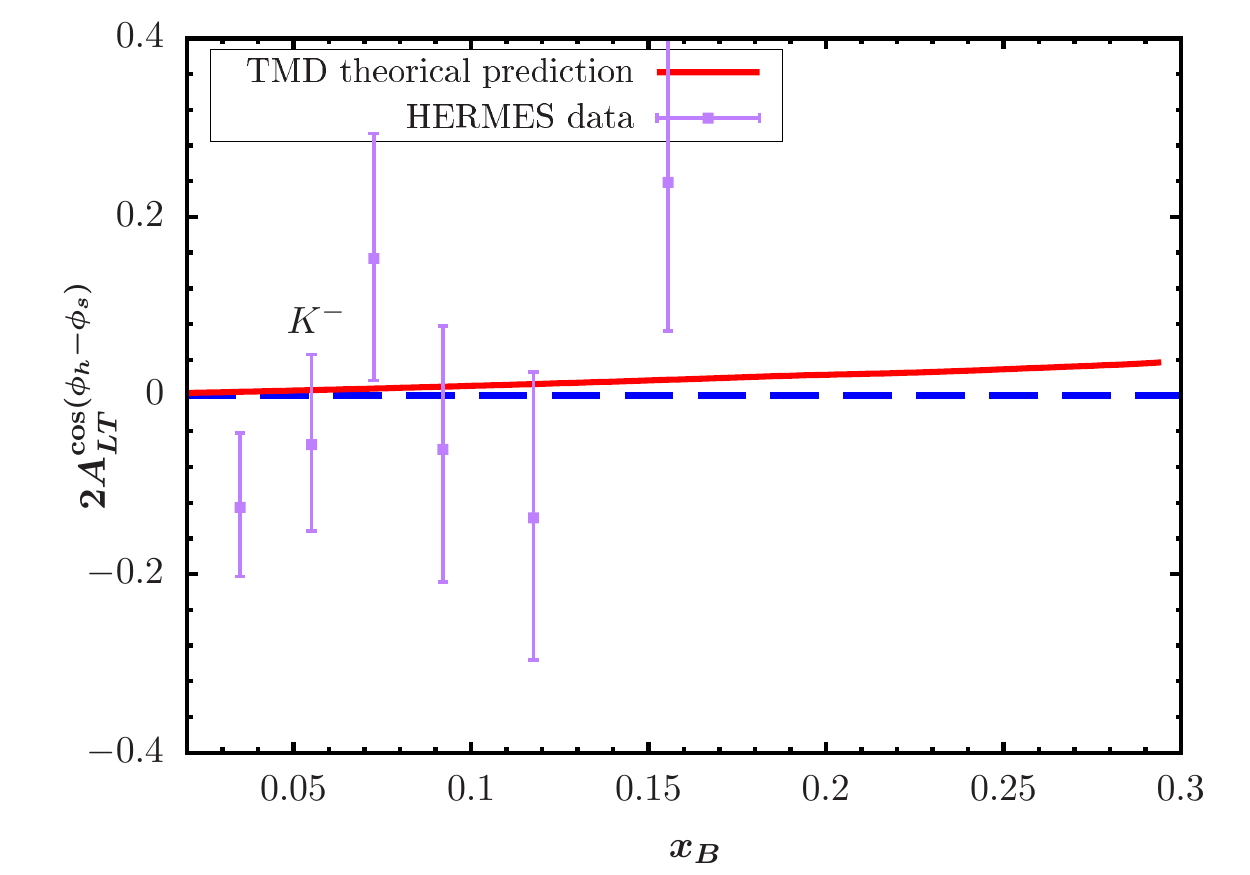}
\includegraphics[scale=0.46]{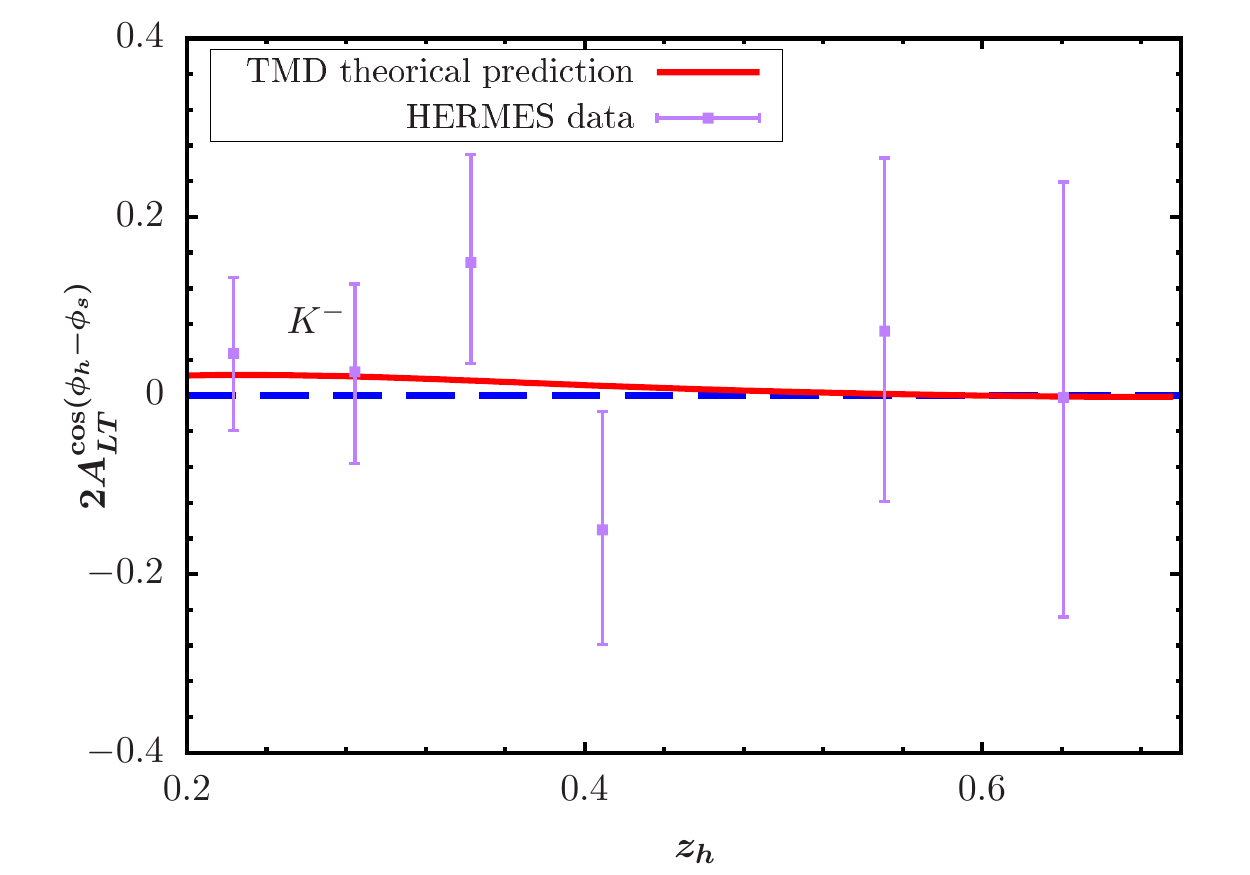}
\includegraphics[scale=0.46]{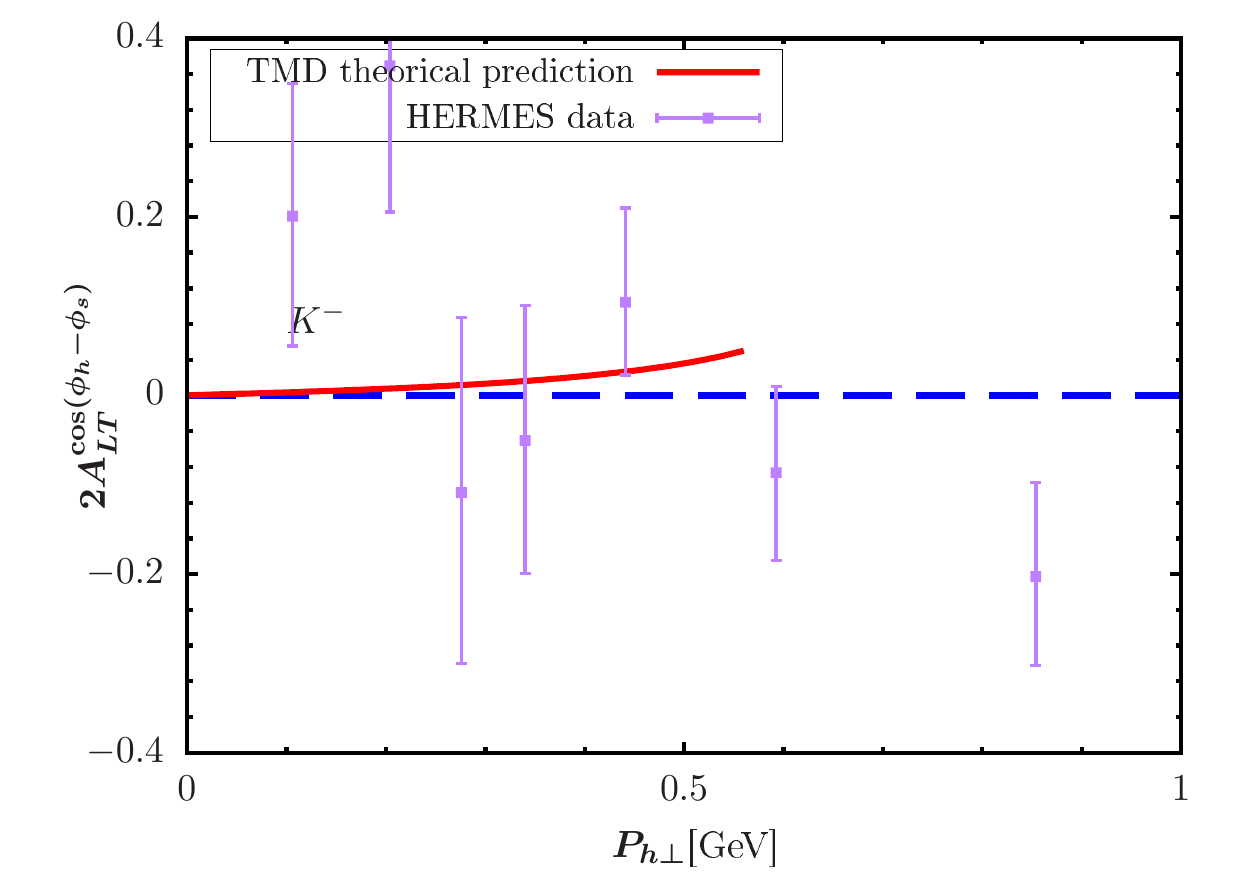}
\caption{ The KM effect calculated within TMD factorization, compared with the HERMES measurement \cite{Airapetian:2009ae} for $K^-$ production.}
\label{fig5}
\end{figure}

\begin{figure}[htp]
\centering
\includegraphics[scale=0.56]{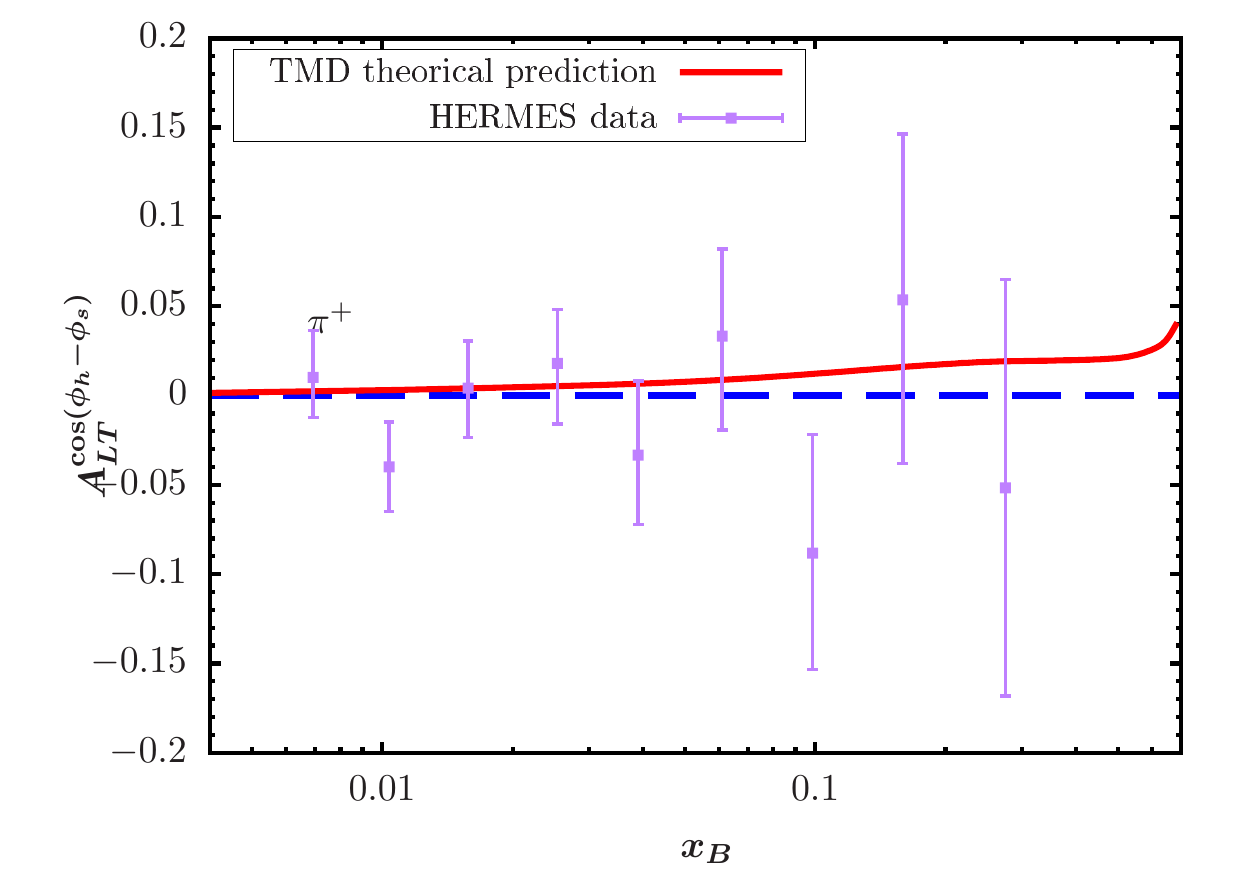}
\includegraphics[scale=0.56]{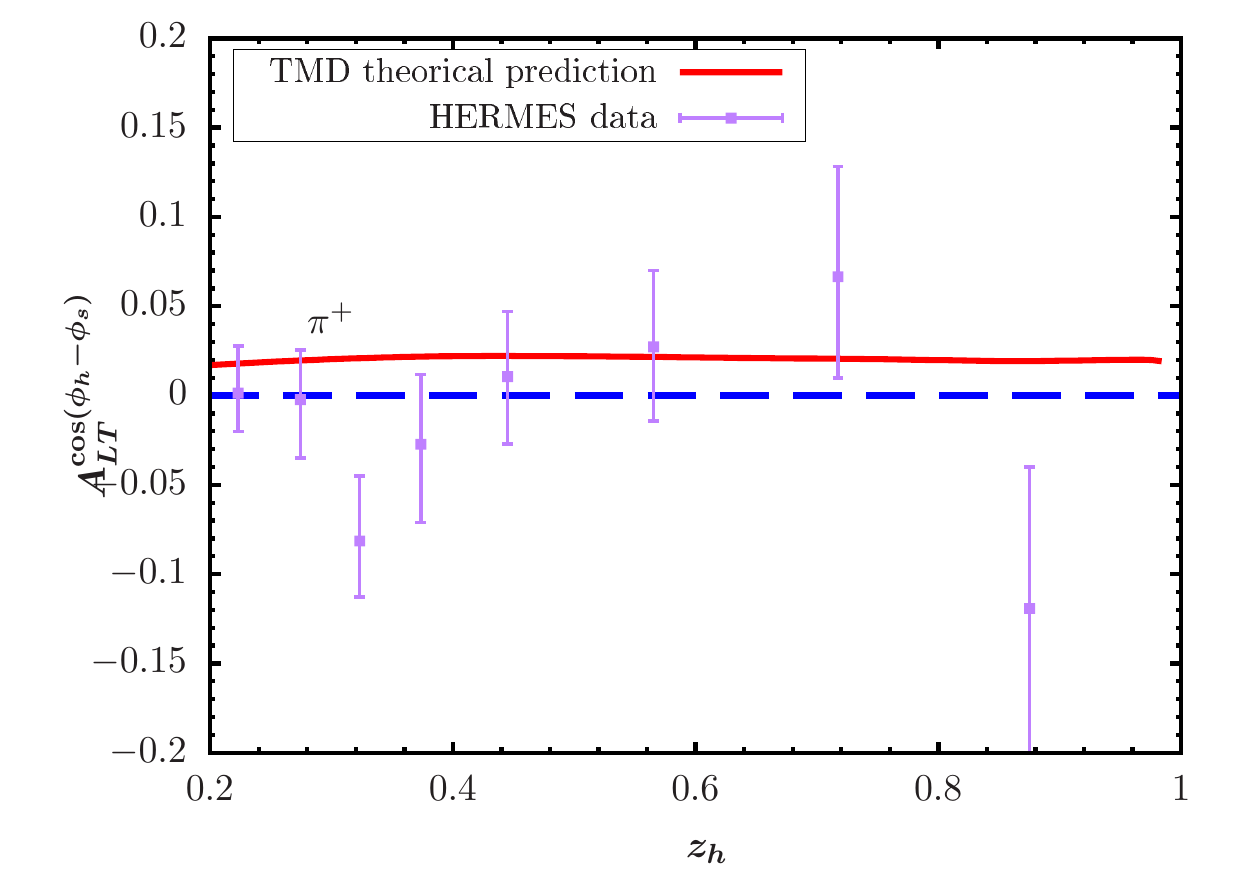}
\\
\includegraphics[scale=0.56]{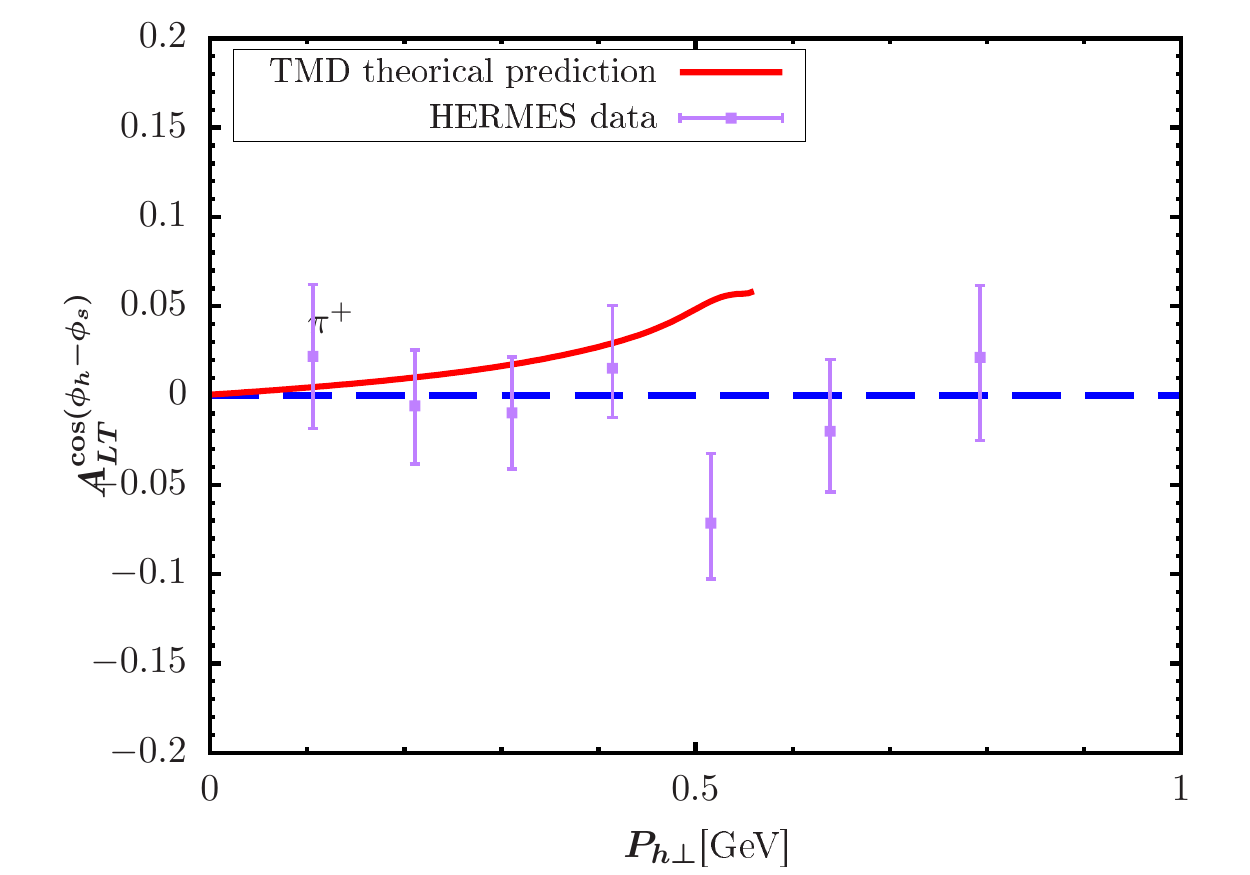}
\includegraphics[scale=0.56]{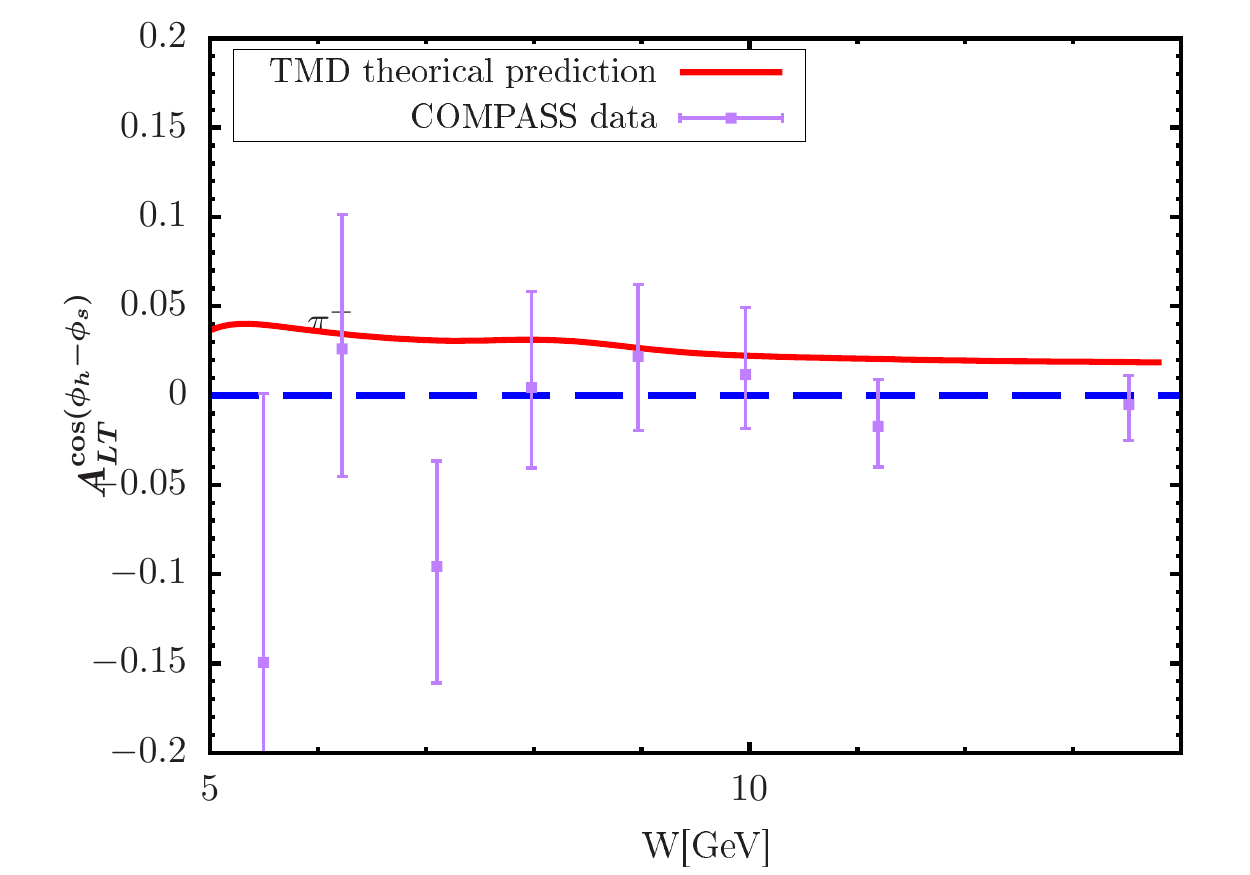}
\caption{ The KM effect calculated within TMD factorization, compared with the COMPASS measurement \cite{Parsamyan:2010se} for $K^-$ production.}
\label{fig6}
\end{figure}

\begin{figure}[htp]
\centering
\includegraphics[scale=0.56]{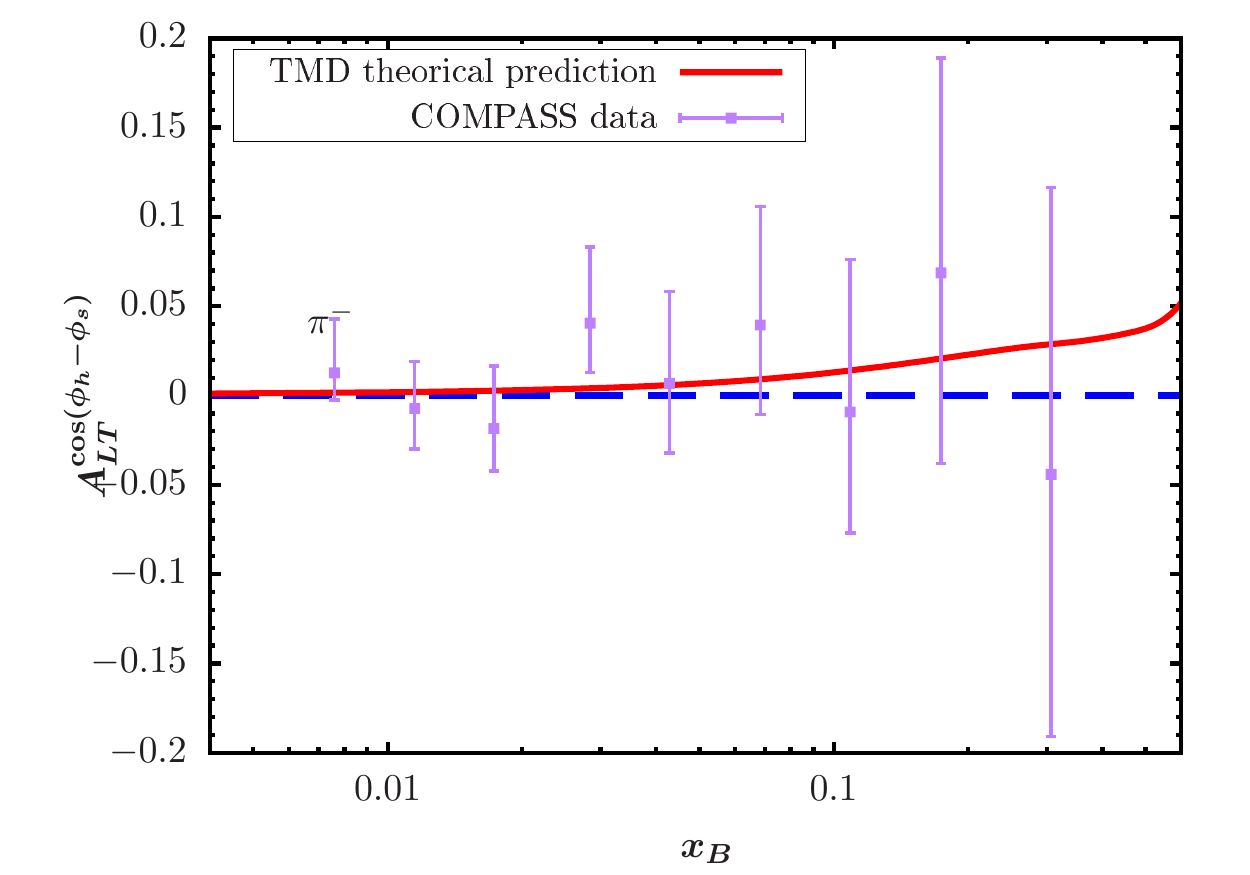}
\includegraphics[scale=0.56]{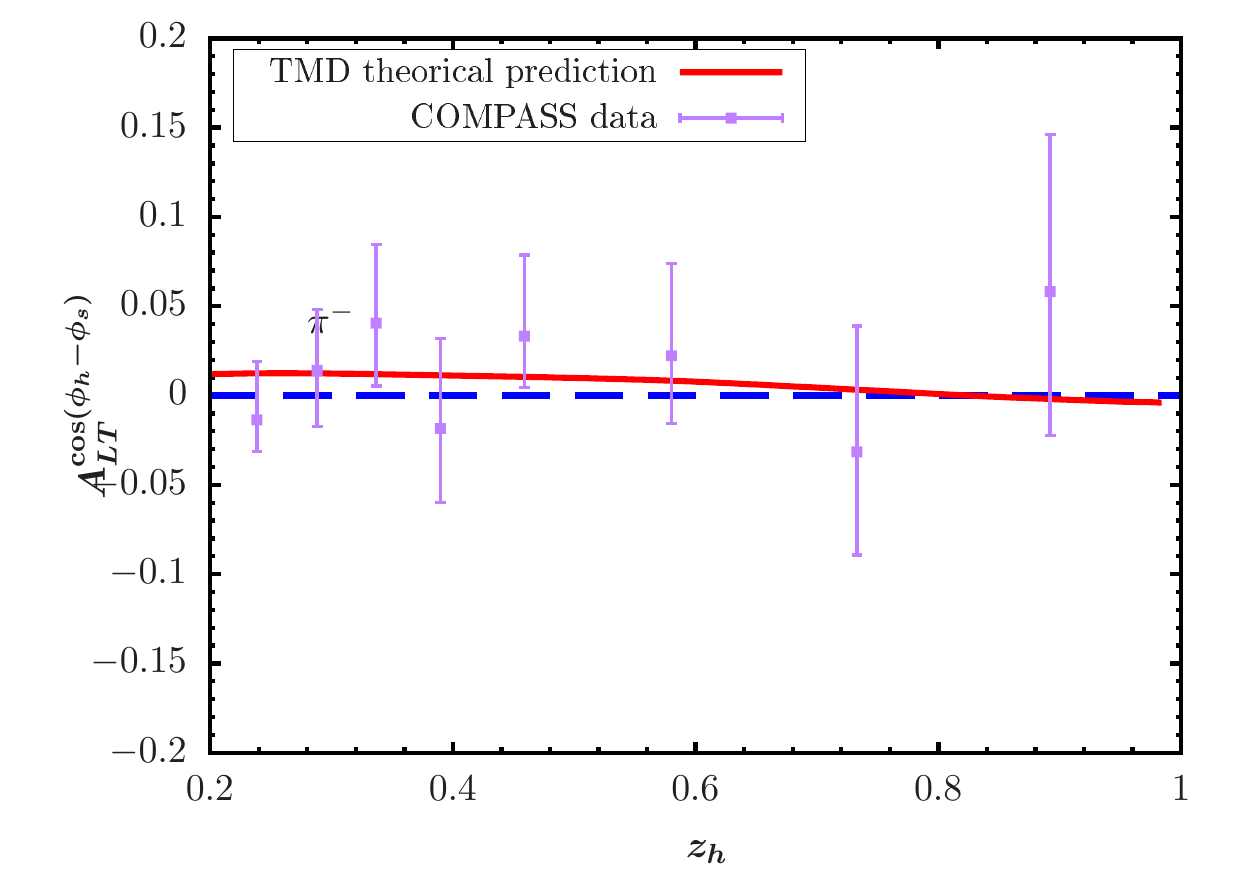}
\\
\includegraphics[scale=0.56]{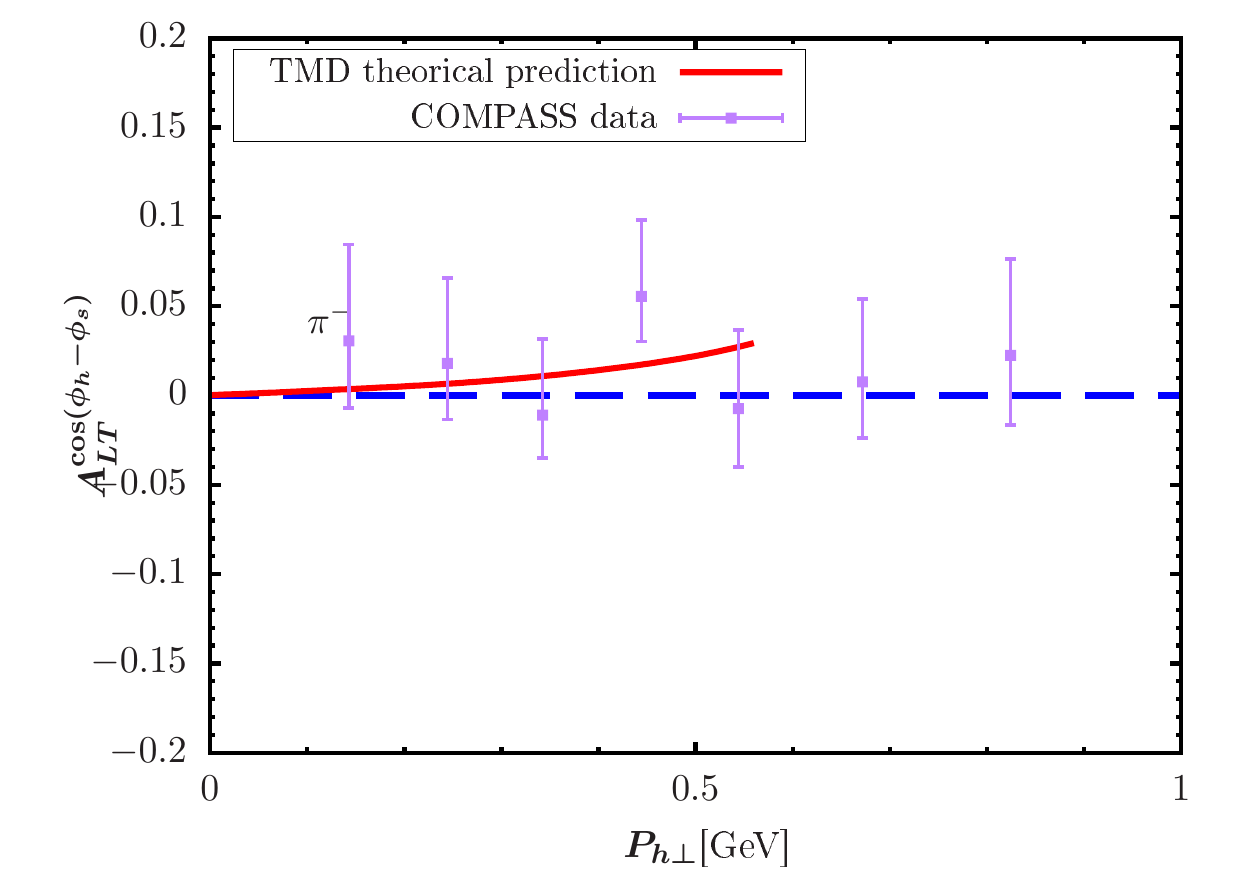}
\includegraphics[scale=0.56]{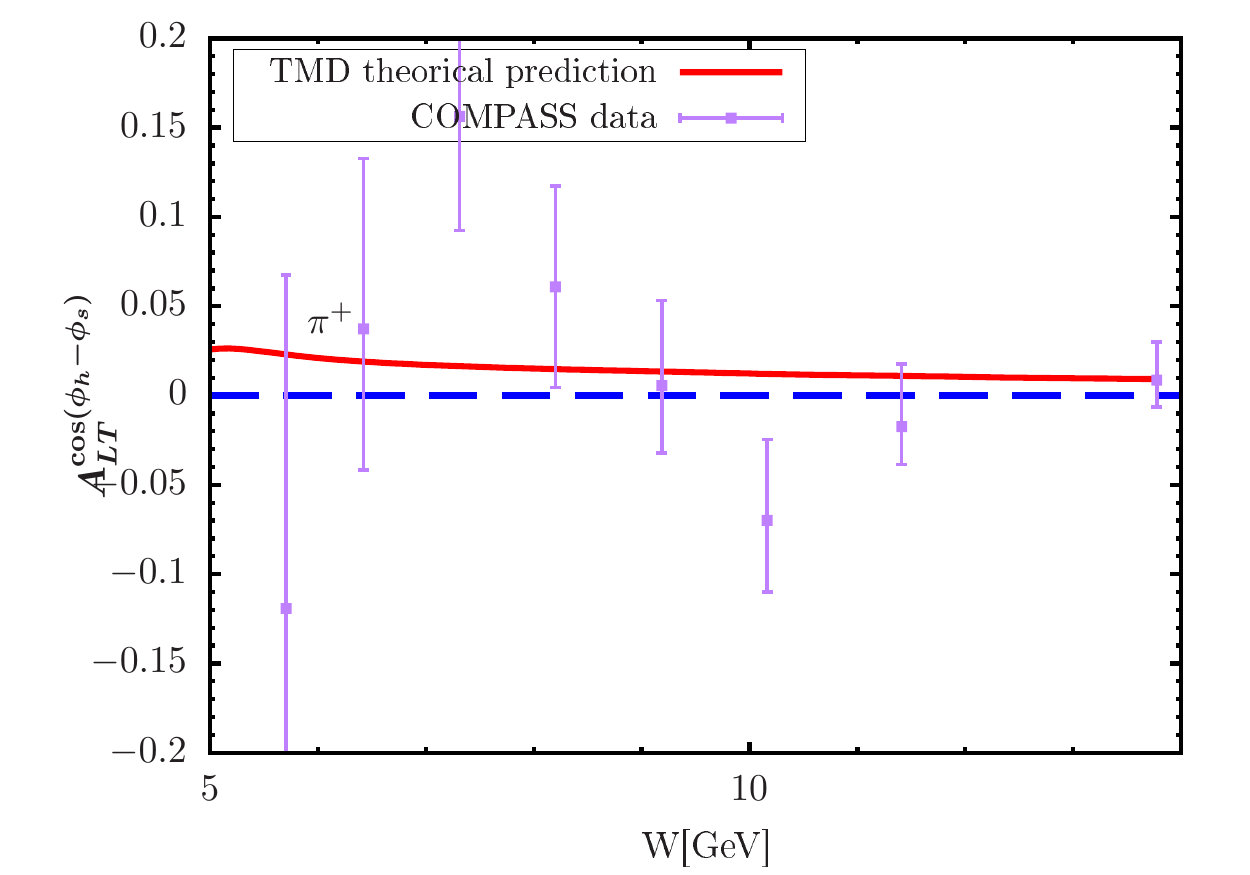}
\caption{ The KM effect calculated within TMD factorization, compared with the COMPASS measurement \cite{Parsamyan:2010se} for $K^-$ production.}
\label{fig7}
\end{figure}

In Figs.\ref{fig1}-\ref{fig7}, we show the results for pion and kaon production.
By integating over the other variables, the $x_B$-, $z_h$- and $P_{h\perp}$-dependent KM effect are depicted in the left, central and right panels of the figure, respectively.  The solid lines represent our model predictions. The full circles with error bars show the preliminary HERMES and COMPASS data for comparison.
For $\pi^-$ and $\pi^0$ production Figs.\ref{fig2}-\ref{fig3} give a good description for the HERMES data, while Fig.\ref{fig1} somewhat overestimate the HERMES data.
For pion production in Figs.\ref{fig1}-\ref{fig3}, the obtained $P_{h\perp}$-dependent effects increase as $P_{h\perp}$ increases, and the largest effect could arrive at 0.15.
As for the $K^+$ production case, the obtained asymmetries in Fig.\ref{fig4} also shows a adjacent result with the HERMES data.
$P_{h\perp}$-dependent effect in Fig.\ref{fig4} can reach nearly 0.1 at point with $x_B = 0.5$.
Fig.\ref{fig5} shows rather small effects for $K^-$ production which is also basically consistent with HERMES data. 
It is desired to mention that when $x_B > 0.5$ the predicted $z_h$-dependent effect nearly becomes zero.
In Figs.\ref{fig6} and \ref{fig7}, we plot the TMD predictions of KM effect for $\pi^+,\pi^-$ production. The cases are similar to the HERMES cases and the predictions are basically consistent with COMPASS data.
The uncertainty may comes from $f_1$,$D_1$,$\tilde{g}$, the evolution kernel of $\tilde{g}$ and the nonperturbative factors. 
From the $g_{1T}^{(1)}$ distributions in Fig.\ref{fig0}, we can find that the TMD evolution effect of $g_{1T}^{(1)}$ distributions is obvious especially in small-$x$ region. In total, the TMD evolution effects are non-negligible and significant in almost all region especially in small-$x$ region.
It is desirable to mention that the data from the HERMES and COMPASS also have large statistical errors. Thus we can not make strong  conclusions from comparing it with our theoretical calculations, but expect more precise data and deeper understanding in SIDIS process.

\section{Conclusion}\label{secIV}

In this paper we have studied the KM effect of a single hadron production in SIDIS within the framework of TMD factorization.
We have applied the energy evolutions of the $\tilde{g}$ function by taking the parametrization at a initial energy $Q_0$ and evolving it to another energy $\mu_b$ through an approximation evolution kernel including only the homogenous terms for the $\tilde{g}$ function. Similarly, the time-like evolution of the unpolarized fragmentation function is also performed by QCDNUM.
Then we reach the $x_B$-, $z_h$- and $P_{h\perp}$-dependent KM effects for the pion and kaon production at the kinematics of HERMES and COMPASS. Then the results are compared with the corresponding HERMES and COMPASS measurements. It is found that the KM effect reached within the TMD factorization and evolution in the corresponding kinematics is basically consistent with the HERMES and COMPASS measurements. Finally we expect more precise data and deeper understanding in SIDIS process, since the data from the HERMES and COMPASS also have large statistical errors.

\begin{acknowledgments}
Hao Sun is supported by the National Natural Science Foundation of China (Grant No.12075043).
\end{acknowledgments}

\bibliographystyle{apsrev4-1}
\bibliography{v1}

\end{document}